\documentclass[11pt]{article}
\pdfoutput=1
\usepackage{jcapmod}
\usepackage{booktabs}
\usepackage[english]{babel}
\usepackage{amsmath,amssymb,amsbsy,amstext, amsthm, simplewick}
\usepackage{graphicx}
\usepackage{amsfonts}
\usepackage{amssymb}
\usepackage{upgreek}
 \usepackage{exscale,relsize}

\allowdisplaybreaks[1]

\usepackage{colortbl}
\definecolor{lightgreen}{cmyk}{0.2, 0, 0.2, 0.2}
\definecolor{lightgray}{cmyk}{0.1,0.2,0,0.1}
\definecolor{lightgray2}{cmyk}{0.1,0.1,0,0.1}

\makeatletter
\newlength{\apb@width}
\newcommand{\autoparbox}[2][c]{\settowidth{\apb@width}{#2}\parbox[#1]{\apb@width}{#2}}
\newcommand{\includegraphicsbox}[2][]{\autoparbox{\includegraphics[#1]{#2}}}
\makeatother

\numberwithin{equation}{section}

\def\beq{\begin{equation}}
\def\eeq{\end{equation}}

\def\bea{\begin{eqnarray}}
\def\eea{\end{eqnarray}}

\def\d{{\rm d}}

\def\beq{\begin{equation}}
\def\eeq{\end{equation}}
\def\bea{\begin{eqnarray}}
\def\eea{\end{eqnarray}}

\def\d{{\rm d}}

\def\d{{\rm d}}

\def\H{{\cal H}}

\def\G{{\cal G}}

\def\0{{\boldsymbol 0}}
\def\q{{\boldsymbol{k}}}
\def\q{{\boldsymbol{q}}}
\def\p{{\boldsymbol{p}}}
\def\v{{\boldsymbol{v}}}
\def\x{{\boldsymbol{x}}}

\def\u{{\boldsymbol{u}}}
\def\D{{\boldsymbol{ \nabla}}}

\def\knl{k_{\mathsmaller{\rm NL}}}
\DeclareRobustCommand{\SkipTocEntry}[4]{}

\newcommand{\vev}[1]{\langle #1 \rangle}

\begin{document}

\begin{titlepage}

\setcounter{page}{1} \baselineskip=15.5pt \thispagestyle{empty}

\bigskip\

\vspace{2cm}
\begin{center}

{\fontsize{20}{28}\selectfont  \sffamily \bfseries Renormalized Halo Bias}

\end{center}

\vspace{0.2cm}

\begin{center}
{\fontsize{13}{30}\selectfont  Valentin Assassi,$^{\bigstar}$ Daniel Baumann,$^{\bigstar}$ Daniel Green,$^{ \blacklozenge, \clubsuit}$ and Matias Zaldarriaga$^{ \spadesuit}$}
\end{center}

\begin{center}

\vskip 8pt
\textsl{$^\bigstar$ D.A.M.T.P., Cambridge University, Cambridge, CB3 0WA, UK}
\vskip 7pt

\textsl{$^ \blacklozenge$ Stanford Institute for Theoretical Physics, Stanford University, Stanford, CA 94305, USA}
\vskip 7pt
\textsl{$^\clubsuit$ Kavli Institute for Particle Astrophysics and Cosmology, Stanford, CA 94305, USA}
\vskip 7pt
\textsl{$^\spadesuit$  Institute for Advanced Study, Princeton, NJ 08540, USA}
\end{center}

\vspace{1.2cm}
\hrule \vspace{0.3cm}
{ \noindent {\sffamily \bfseries Abstract} \\[0.1cm]
This paper provides a systematic study of renormalization in models of halo biasing.
Building on work of McDonald, we show that Eulerian biasing is only consistent with renormalization if non-local terms and higher-derivative contributions are included in the biasing model.  We explicitly determine the complete list of required bias parameters for Gaussian initial conditions, up to quartic order in the dark matter density contrast and at leading order in derivatives.  At quadratic order, this means including the gravitational tidal tensor, while at cubic order the velocity potential appears as an independent degree of freedom.
 Our study naturally leads to an effective theory of biasing in which the halo density is written as a double expansion in fluctuations and spatial derivatives.  
 We show that the bias expansion can be organized in terms of Galileon operators which aren't renormalized at leading order in derivatives.  
Finally, we discuss how the renormalized bias parameters impact the statistics of halos. 

\noindent}

\hrule

\vspace{0.6cm}

\end{titlepage}

 \tableofcontents

\newpage

\section{Introduction}

\begin{figure}[h!]
   \centering
       \includegraphics[scale =0.8]{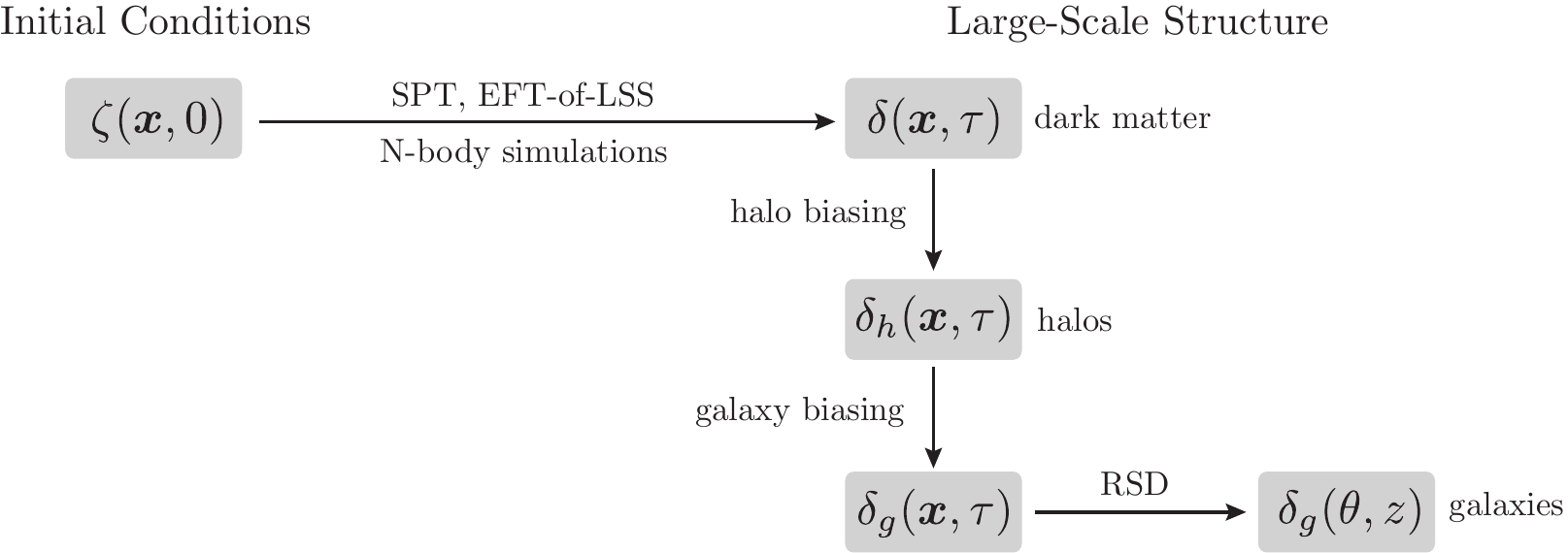}
    \caption{Non-linearities in the gravitational evolution, in the biasing and in redshift space distortions~(RSD) complicate the relationship between the primordial initial conditions and large-scale structure observables.}
  \label{fig:schematic}
\end{figure}

Over the next decade, large-scale structure (LSS) surveys will play an increasingly important role in the measurement of cosmological parameters and as a probe of initial conditions.  
In order to relate 
late-time observables 
to the physics of the early universe,
several sources of secondary non-linearities need to be understood (see fig.~\ref{fig:schematic}).  
Reducing the theory error is essential if the full potential of future surveys is to be realized.\footnote{The number of useful modes in galaxy surveys scales as the cube of the maximum wavenumber, $k_{\rm max}$, at which the theoretical predictions can still be trusted.  Even a relatively modest gain in $k_{\rm max}$ can therefore dramatically impact the scientific potential of galaxy surveys (but see~\cite{Rimes:2005xs, Crocce:2005xz}).}  
Non-linearities in the gravitational evolution can be characterized by numerical N-body simulations~\cite{Springel:2005mi} and, on sufficiently large scales, by perturbation theory~\cite{Bernardeau:2001qr, Bernardeau:2013oda}.
Less well understood are non-linearities in the {\it biasing} between the clustering of galaxies and the underlying dark matter density. 

\vspace{0.2cm}

\vskip 4pt
The biasing problem is already visible in dark matter-only simulations, where it is reflected in the biasing of dark matter halos.
On large scales, linear biasing has been shown to be a
good approximation:
\beq
\delta_{h} = b_1 \delta  \ ,
\eeq
where $\delta_h$ and $\delta$ are the density contrasts of the halos and the dark matter, respectively, and the bias parameter $b_1$ is an unknown coefficient (to be fit to data).  However, linear biasing is known to fail on small scales where non-linearities becomes important.   
One common procedure for describing halos beyond the linear biasing model is {\it local Eulerian biasing}~\cite{Fry:1992vr} which assumes that the halo density contrast is a local function of the dark matter density, $\delta_h(\x,\tau)={\cal F}[\delta(\x, \tau)]$. Formally, we might write this relation as a Taylor expansion
\beq
\delta_h(\x, \tau) = \sum_{n=0}^\infty  \frac{b_n^{(0)}}{n!} \, \delta^n(\x, \tau)\label{equ:Fry} \ .
\eeq
Local biasing is motivated both as a natural generalization of linear biasing and as a consequence of a number of semi-analytic models of halo formation.  
It is also often employed in data analysis~\cite{Gaztanaga:1994us, Scoccimarro:2000sp, Feldman:2000vk, Verde:2001sf, Gaztanaga:2005an, McBride:2010zp}.
However, the meaning of (\ref{equ:Fry}) is far from clear, as we need to define $\delta^n(\x, \tau)$ for $n>1$.  
In particular, non-linear quantities like $\delta^n$ receive contributions from all scales and are therefore not necessarily small, even on large scales. 
A common procedure is to smooth $\delta$ with a window function that removes power below some length scale~$\Lambda^{-1}$.  However, this only masks the problem.
The scale~$\Lambda^{-1}$ is arbitrary and should not appear in physical quantities.\footnote{Semi-analytic models only add to the confusion, as they often identify $\Lambda^{-1}$ with the Lagrangian size of the halo, which is a physical scale.}  
In~\cite{McDonald:2006mx}, McDonald showed how to reorganize the bias expansion in a way that makes this property manifest (see also~\cite{McDonald:2009dh, Schmidt:2012ys}). 
In this paper, we revisit his idea of {\it renormalized halo bias}.

\vskip 4pt
The key feature of renormalization is that, although $\delta^n$ may be large (or depend on an unphysical smoothing scale $\Lambda$), all large (or $\Lambda$-dependent) contributions can be systematically removed by adding local counterterms 
\beq
[\delta^n](\x, \tau) \equiv \delta^n(\x, \tau) + \sum_{\widetilde {\cal O}} Z^{(\delta^n)}_{\widetilde{\cal O}} \hskip 1pt\widetilde{\cal O}(\x, \tau)  \, \ll\, 1\ .
\eeq  
We will find that consistent renormalization requires additional fields $\widetilde {\cal O}$ beyond those appearing in the local Eulerian biasing model~(\ref{equ:Fry}). For instance, at quadratic order, we have to add the tidal tensor~$(\nabla_i \nabla_j \Phi_g)^2$~\cite{McDonald:2009dh, Chan:2012jj, Baldauf:2012hs}, where $\Phi_g$ is the gravitational potential, while, at cubic order, the velocity potential $\Phi_v$ has to be introduced as an independent degree of freedom.  In addition, higher-derivative terms must be included to remove the sub-leading $\Lambda$-dependence.  Finally, no stochastic parameters are needed to remove divergences in correlation functions at separated points, although one may choose to include them to remove delta-function localized divergences.  In general, we find that all the terms that are allowed by symmetry are required in order to define the renormalized fields~$[{\cal O}] = \{ [\delta^n],[\widetilde{\cal O}]\}$.\footnote{In this paper, we will only consider the case of Gaussian initial conditions. Additional terms can arise for non-Gaussian initial conditions~\cite{McDonald:2008sc}. We will discuss this elsewhere~\cite{NG-progress}.} 

Renormalized halo biasing is naturally an {\it effective theory},
 in which the halo density contrast is written as a double expansion in terms of powers of the fluctuations and their derivatives.  In order to be well-defined, this expansion 
has to be organized in terms of the renormalized quantities:
\beq
\delta_h(\x, \tau) \, =\,  \sum_{\cal O} b_{{\cal O}}^{(R)} \, [{\cal O}](\x, \tau)\ ,\label{equ:dh}
\eeq
where $b_{{\cal O}}^{(R)}$ are the {\it renormalized bias parameters} which by construction do {\it not} depend on the smoothing of the density field.
It is to be expected that the biasing model (\ref{equ:dh}) will contain all terms~${\cal O}$ consistent with the symmetries of the dark matter equations of motion, as was emphasized by McDonald and Roy~\cite{McDonald:2009dh}~(see also \cite{Kehagias:2013rpa}).\footnote{The form of~(\ref{equ:dh}) was also argued to arise from the non-linear time evolution of local Lagrangian biasing~\cite{Chan:2012jj}.}
After renormalization, terms with higher number of fields and higher number of spatial derivatives are suppressed.  To describe the halo statistics to a finite precision then only requires a finite number of terms in (\ref{equ:dh}).
How many terms need to be retained
  depends on the power spectrum of the initial conditions in the same way as in the effective field theory of large-scale structure (EFT-of-LSS)~\cite{Carrasco:2012cv, Pajer:2013jj, Carrasco:2013mua}.  
  In principle, the different terms in the bias expansion can be distinguished by measuring correlation functions of the dark matter halos and fitting for the bias parameters $b_{{\cal O}}^{(R)}$.  For the gravitational tidal tensor this has been demonstrated in \cite{Baldauf:2012hs, Sheth:2012fc}. 
Here, we lay out the basic steps towards a systematic treatment of the effective theory of halo biasing.

\vskip 6pt
The outline of the paper is as follows: In Section~\ref{sec:Renorm}, we give a pedagogical introduction to renormalization in structure formation. We present the renormalization of $\delta^2$ as an explicit example. In Section~\ref{sec:Renorm2}, we generalize this to all terms in the local Eulerian biasing model.  We classify the terms that have to be added to the biasing model to make it consistent with renormalization.
In Section~\ref{sec:HaloStatistics}, we show how these terms affect halo statistics at one-loop order. 
Our conclusions appear in Section~\ref{sec:Conclusions}.
Four appendices contain technical details: In Appendix~\ref{sec:mixing},
we discuss how the renormalized quantities depend on the renormalization scale.
In Appendix~\ref{ap:HOO}, we present the renormalization of $\delta^n$, for $n > 2$.
In Appendix~\ref{app:Galileon}, we prove a non-renormalization theorem for Galileon operators. In Appendix~\ref{app:Bispectrum}, we give the details of the one-loop halo bispectrum.

\vskip 6pt
We will use the following notation and conventions:
Conformal time is denoted by $\tau$ and comoving coordinates by $\x$.
Overdots denote derivatives with respect to conformal time and $\H \equiv \dot a/a$ is the comoving Hubble parameter. For momentum integrals, we use the shorthand $\int_\p \equiv \int \d^3 p/(2\pi)^3$. 
We define the normalized inner product of two vectors as $\mu_{\p,\q}\equiv \p\cdot\q/pq$ and sometimes use $\sigma^2_{\p,\q} \equiv \mu^2_{\p,\q} - 1$.
A prime on a correlation function, $\langle \ldots \rangle'$, indicates that the overall delta function is dropped.  We will use $P(q) \equiv \langle \delta^{(1)}_\q \delta^{(1)}_{\q'}\rangle'$ for the linear power spectrum (which is often called $P_{11}$) and $P_{mm}(q) \equiv \langle \delta_\q \delta_{\q'}\rangle'$ for the non-linear power spectrum.  Sometimes, we write
 $P_i \equiv P(q_i)$. Renormalized operators are denoted by square brackets, $[{\cal O}]$.  We use two definitions for the bias parameters of $\delta^n$, namely $b_n$ and $b_{\delta^n}$, which are related by $b_{\delta^n}\equiv b_n/n!$. When we present numerical results, the linear power spectrum is computed with CAMB~\cite{Lewis:1999bs}, using a $\Lambda$CDM cosmology with the standard cosmological parameters, $\Omega_{\Lambda} = 0.73$, $\Omega_b h^2 =0.02$, $\Omega_m h^2 = 0.12$, $h = 0.7$, and power law initial conditions, with $n_s=0.96$ and $A_s = 2.2 \times 10^{-9}$.

\section{Renormalization in Structure Formation}
\label{sec:Renorm}

Before explaining the renormalization of local Eulerian biasing, we first review, in \S\ref{sec:RDM}, dark matter perturbation theory in light of the recently developed EFT-of-LSS~\cite{Baumann:2010tm, Carrasco:2012cv} (see also~\cite{Carrasco:2013mua, Carrasco:2013sva, Porto:2013qua, Carroll:2013oxa, Mercolli:2013bsa, Pajer:2013jj, Hertzberg:2012qn}). 
We then discuss, in~\S\ref{sec:RHB}, the renormalization of the simplest non-linear term in the local Eulerian biasing model, namely $\delta^2$.
The complete renormalization and further technical details will be presented in Section~\ref{sec:Renorm2} and in the appendices.

\subsection{Renormalized Dark Matter} 
\label{sec:RDM}

\subsubsection{Fluid Equations}
\label{sec:fluideqs}

On large scales, dark matter acts as a pressureless fluid.  In the Newtonian approximation, the dark matter density contrast $\delta \equiv\delta \rho/\bar\rho$ and the dark matter velocity $\v$ satisfy the following evolution equations 
\begin{align}
\left( \partial_\tau +\v\cdot\D\right)\delta &\, =\, -(1+\delta)\D\cdot\v\ ,\label{eq:cont}\\
\left(\partial_\tau +\v\cdot\D\right)\v &\,=\, -\H\v-\D\Phi_g\ ,\label{eq:Euler}
\end{align}
where $\Phi_g$ is the gravitational potential.   
 In a matter-dominated universe, $\Phi_g$ is sourced by $\delta$
\beq
\nabla^2\Phi_g=\frac{3}{2}\H^2 \Omega_m\delta\ ,
\eeq
where $\Omega_m = 1$ in an Einstein-de Sitter (EdS) universe.
If the velocity field is irrotational, it is described completely by its divergence 
\beq
\theta \equiv \D \cdot \v \equiv \nabla^2 \Phi_v\ ,
\eeq 
where $\Phi_v$ is the velocity potential.  We will use $\Phi_g$ and $\Phi_v$ as our fundamental fields.

As a consequence of the {\it equivalence principle}, the evolution equations are invariant under the {\it boost symmetry}
\beq
 \x\ \mapsto\, \x\, -\ T\,\u \quad , \quad \v\, \mapsto\, \v - \dot T\,\u\ , \label{equ:trans}
 \eeq
where $T(\tau)$ is an arbitrary function of the conformal time $\tau$ and $\u$ is the boost velocity. 
The boost transformation (\ref{equ:trans}) shifts the gradient of the gravitational potential and the velocity potential,
\begin{align}
\Phi_g&\, \mapsto\, \Phi_g\, +\, \big[\H\dot T+\ddot T\big] \u \cdot\x\ , \label{equ:Phig}\\\Phi_v&\, \mapsto\, \Phi_v\, -\, \dot T\hskip 2pt \u \cdot\x\ .
\end{align}
Finally, in EdS, the fluid equations are invariant under a {\it Lifshitz symmetry} 
\beq
\tau \mapsto \lambda^z \tau\quad , \quad \x \mapsto \lambda \x \ , 
\eeq
for a generic weight $z$. The potentials transform as
\begin{align}
\Phi_g&\, \mapsto\, \lambda^{2(1-z)} \Phi_g\ ,\\
\Phi_v&\, \mapsto\, \lambda^{2-z} \Phi_v\ .
\end{align}
For power law initial conditions, $\vev{\delta_\q(\tau_{\rm in})\,\delta_{\q'}(\tau_{\rm in})}' \propto q^n$, the rescaled solutions have the same initial conditions iff $4z=n+3$.

Further discussion of the symmetries of the Newtonian fluid equations can be found in~\cite{Peloso:2013zw, Kehagias:2013yd}.
Conserved tracers of the dark matter density satisfy the same equations and the same symmetries as the dark matter perturbations.  Of course, in reality, the number of halos is not conserved, so the evolution equations will contain extra source terms.  
However, we will only require that the halo density contrast $\delta_h$ is a scalar under the transformation (\ref{equ:trans}).  Any equation governing the time evolution of halos will only impact the time evolution of the bias parameters which is beyond the scope of this work.

\subsubsection{Standard Perturbation Theory}

We briefly review standard perturbation theory (SPT) in an Einstein-de Sitter universe~\cite{Bernardeau:2001qr}. 
It is convenient to write the equations of motion (\ref{eq:cont}) and (\ref{eq:Euler}) in Fourier space
\begin{align}
 \dot \delta + \theta &\,=\, - \theta \star \delta\ ,  \label{equ:d1}\\
\dot \theta + \H \theta + \frac{3}{2}\H^2 \Omega_m \delta &\, =\, - \theta \star \theta\ , \label{equ:t1}
\end{align}
where the left-hand sides capture the linear evolution and the right-hand sides contain non-linear  convolutions
\begin{align}
[ \theta \star \delta]_\q &\,\equiv\, \int_{\q_1}  \alpha(\q,\q_1)\, \theta_{\q_1} \delta_{\q - \q_1}\ ,  \qquad \alpha(\q,\q_1) \equiv \frac{\q \cdot \q_1}{q_1^2} \ , \\
[ \theta \star \theta]_\q &\,\equiv\, \int_{\q_1} \, \beta(\q,\q_1,\q - \q_1)\, \theta_{\q_1} \delta_{\q- \q_1} \ , \quad \beta(\q,\q_1,\q_2) \equiv \frac{q^2 (\q_1 \cdot \q_2)}{2 q_1^2 q_2^2} \ .
\end{align}
Eqs.~(\ref{equ:d1}) and (\ref{equ:t1}) can be solved once the initial conditions for $\delta$ and $\theta$ have been specified. At sufficiently early times, the dark matter density contrast and the velocity divergence are small. Consequently, the equations of motion can be solved order by order in the initial conditions. In an Einstein-de Sitter universe, the solution can formally be written as\hskip 1pt\footnote{Although strictly speaking this form of the result is only valid for the EdS universe, it can approximately be extended to arbitrary values of $\Omega_m$ and $\Omega_\Lambda$, by replacing the EdS growth factor $a$ by the growth factor of the corresponding cosmology $D(a)$~\cite{Bernardeau:2001qr}. }
\beq
\delta(\x,\tau)\, =\, \sum_{n=1}^\infty a^{n}(\tau) \hskip 1pt \delta^{{(n)}}(\x, \tau_{\rm in})\qquad{\rm and}\qquad\theta(\x,\tau)\, =\,  -\H(\tau)\sum_{n=1}^\infty a^{n}(\tau) \hskip 1pt \theta^{{(n)}}(\x, \tau_{\rm in})\ ,\label{eq:deltaSPT}
\eeq
where only the growing mode has been kept.\footnote{Formally, this is equivalent to imposing the initial conditions at $\tau_{\rm in}\to0$.}  In Fourier space, the fields $\delta^{{(n)}}$ and $\theta^{{(n)}}$ can be written as
\begin{align}
\delta^{{(n)}}_\q(\tau_{\rm in})&\, =\,  \int_{\q_1}\cdots\int_{\q_n} (2\pi)^3\delta_D(\q_1+\cdots+\q_n-\q)\, F_n(\q_1,\cdots,\q_n)\,\delta^{{(1)}}_{\q_1}(\tau_{\rm in})\cdots\delta^{{(1)}}_{\q_n}(\tau_{\rm in})\ ,\label{eq:deltan}\\
\theta^{{(n)}}_\q(\tau_{\rm in})&\, =\, \int_{\q_1}\cdots\int_{\q_n}(2\pi)^3\delta_D(\q_1+\cdots+\q_n-\q)\, G_n(\q_1,\cdots,\q_n)\,\delta^{{(1)}}_{\q_1}(\tau_{\rm in})\cdots\delta^{{(1)}}_{\q_n}(\tau_{\rm in})\ ,
\end{align}
where the kernel functions $F_n$ and $G_n$ can be computed iteratively~\cite{Bernardeau:2001qr}. In particular, the first-order kernels are just $F_1=G_1=1$, while the second-order kernel are given by\footnote{We will use the {\it symmetrized} kernels, obtained by summing over all permutations of the momenta.}
\begin{align}
F_2(\q_1,\q_2)&=\frac{5}{7} + \frac{\mu_{12}}{2}\left(\frac{q_1}{q_2}+\frac{q_2}{q_1}\right)+\frac{2}{7}\mu_{12}^2\ , \label{equ:F2} \\
G_2(\q_1,\q_2)&=\frac{3}{7} + \frac{\mu_{12}}{2}\left(\frac{q_1}{q_2}+\frac{q_2}{q_1}\right)+\frac{4}{7}\mu_{12}^2\ , \label{equ:G2}
\end{align}
with $\mu_{12}\equiv\q_1\hskip -1pt\cdot\hskip -1pt\q_2/q_1q_2$. 
The initial conditions are encoded in the linear dark matter density contrast $\delta^{{(1)}}_\q(\tau_{\rm in})$. We will assume that the initial conditions are Gaussian,\footnote{We will treat the non-Gaussian case in~\cite{NG-progress}.} so that the statistics of the initial fluctuations $\delta^{{(1)}}$ is completely determined by its power spectrum
\beq
\vev{\delta^{{(1)}}_\q(\tau_{\rm in})\,\delta^{{(1)}}_{\p}(\tau_{\rm in})}=P(q,\tau_{\rm in})\, (2\pi)^3\delta(\q + \p)\ .
\eeq

\vskip 4pt
The computation of correlation functions can be organized using Feynman diagrams. Each field $\delta^{{(n)}}$ is represented by a vertex with $n$ external legs $\delta^{(1)}$:
\beq
\delta^{{(n)}}_\q \quad = \quad \includegraphicsbox[scale=1.2]{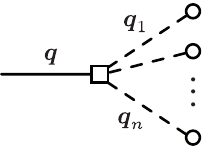}\qquad =\qquad F_n(\q_1,\cdots,\q_n)\, (2\pi)^3\delta_D(\q_1+\cdots+\q_n-\q)\ .  
\eeq
To compute an $N$-point correlation function in an {EdS universe} with {Gaussian initial conditions}, we use the following Feynman rules:   
\begin{enumerate}
\item Draw every connected graph with $N$ vertices.
\item To each vertex with $n$ external legs (with outgoing momenta $\q_i$) assign the factor
\beq n!\, F_n(\q_1,\cdots,\q_n)\, (2\pi)^3\delta_D(\q_1+\cdots+\q_n-\q)\ ,\eeq
where $n!$ corresponds to the symmetry factor.
\item To each propagator assign the {\it time-evolved} linear power spectrum,\footnote{For simplicity, we will often suppress the time coordinate and write $P(q) \equiv P(q,\tau)$. } 
\beq
P(q,\tau)\equiv \vev{\delta^{(1)}_{\q}(\tau)\delta^{(1)}_{\q'}(\tau)}'=a^2(\tau)P(q,\tau_{\rm in})\ ,
\eeq
where $q$ is the magnitude of the momentum flowing in the propagator.

\item Integrate over each loop with measure of integration 
\beq
\int_\q \, \equiv\, \int\frac{\d^3 q}{(2\pi)^3}\ ,
\eeq 
taking into account the symmetry factor of the loop.
\end{enumerate}

\subsubsection{Effective Theory of Large-Scale Structure}
\label{sec:EFTofLSS}

As an example, we compute the dark matter power spectrum $\langle \delta_\q \delta_{\q'} \rangle' \equiv P_{mm}(q)$ at one-loop order, i.e.~at fourth order in the initial conditions $\delta^{(1)}$: 
\begin{align}
P_{mm}(q) &\quad =\quad \ \ \ \, P_{11}(q) \qquad  + \qquad  \ \ \ P_{13}(q) \qquad   \quad + \qquad \, P_{22}(q) \quad \, \ \ \, ,\\
&\quad =\quad \includegraphicsbox[scale=1]{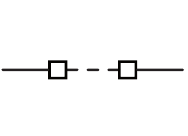} \quad+\quad2\times \includegraphicsbox[scale=1]{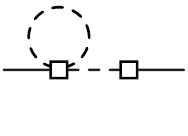}\quad+\quad\includegraphicsbox[scale=1]{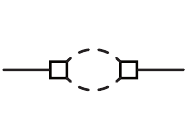} \quad ,
\end{align}
where $P_{nm} \equiv \langle \delta^{(n)}\delta^{(m)} \rangle' $ and $P_{11} \equiv P$.
The first one-loop contribution, $P_{13}$, has the following explicit form
\beq
P_{13}(q) = 3P(q) \int_{\p}F_{3}(\p,-\p,\q)P(p)\ .
\eeq
This is an integral over {\it all} comoving wavenumbers $p$, including those for which perturbations have already grown non-linear and are therefore outside the regime of validity of SPT. Consequently, one cannot take this first-order correction seriously, regardless of whether this integral diverges or not. To make sense of perturbation theory at the loop level, an effective field theory~(EFT) approach has recently been developed~~\cite{Baumann:2010tm, Carrasco:2012cv} (see also~\cite{Carrasco:2013mua, Carrasco:2013sva, Porto:2013qua, Carroll:2013oxa, Mercolli:2013bsa, Pajer:2013jj, Hertzberg:2012qn}).  This approach describes the dark matter on scales larger than some cut-off scale $\Lambda^{-1}\gtrsim k_{\mathsmaller{\rm NL}}^{-1}$, while systematically keeping track of the effects of short-distance fluctuations on long-wavelength observables through modifications to the Euler equation~(\ref{eq:Euler}).  
These new ``fluid" equations modify the SPT solution~$\delta_{\mathsmaller{\rm SPT}}$ to
\beq
\delta\, =\, \delta_{\mathsmaller{\rm SPT}}\, +\, \tilde\delta\ ,\label{eq:deltarenorm}
\eeq
and similarly for the velocity divergence $\theta$. 
Here, $\tilde\delta$ is the new solution generated by the new parameters in the fluid equations. This new solution can also be written as an expansion in the initial conditions \cite{Carrasco:2013mua}
\beq
\tilde\delta(\x, \tau)=\sum_{n=0}^\infty\ a^{n+2}(\tau)\,\tilde\delta^{(n)}(\x, \tau_{\rm in})\ ,
\eeq
with
\beq
\tilde\delta^{(n)}_{\q}(\tau_{\rm in})=\int_{\q_1}\cdots\int_{\q_n} (2\pi)^3\delta_D(\q_1+\cdots+\q_n-\q)\,\tilde F_n(\q_1,\cdots,\q_n; \Lambda)\, \delta^{(1)}_{\q_1}(\tau_{\rm in})\cdots\delta^{(1)}_{\q_n}(\tau_{\rm in})\ .
\eeq
The new kernel functions $\tilde F_n$ depend explicitly on the cut-off $\Lambda$ in such a way as to cancel the cut-off dependence arising from loop diagrams. We will use the following diagrammatic representation
\beq
\tilde\delta^{(n)}_\q  \quad = \quad  \includegraphicsbox[scale=1.2]{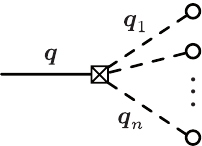}  \quad = \quad  \tilde F_n(\q_1,\cdots,\q_n; \Lambda)\,(2\pi)^3\delta_D(\q_1+\cdots+\q_n-\q)\ . \label{equ:EFTCT} 
\eeq
The renormalized one-loop power spectrum can be written diagrammatically as
\beq
P_{mm}(q)\ =\ \includegraphicsbox[scale=1]{P11}\ +\ \ 2\times\Big(\includegraphicsbox[scale=1]{P13}\ +\ \includegraphicsbox[scale=1]{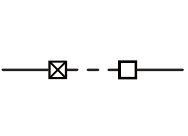}\Big)\ +\ \includegraphicsbox[scale=1]{P22}\ , \label{equ:PNL}
\eeq
where a counterterm proportional to $\tilde\delta^{(1)}$ was added to cancel the cut-off dependence coming from~$P_{13}$.

The second one-loop contribution, $P_{22}$, can be written as the sum of a finite ($\Lambda$-independent) part and a divergent ($\Lambda$-dependent) part. However, the divergent part is proportional to integer powers of~$q^2$ which correspond to delta-function localized contributions in position space. Such {\it contact terms} vanish when the two-point function is evaluated at separated points.  
We will ignore these contact terms 
and therefore our formulas will be correct\footnote{For correlation functions of conserved quantities like $\delta$, some contact terms can be forbidden.  We will not assume any conservation law for halos and therefore their correlation functions may include all possible contact terms.} up to terms analytic in the momenta.\footnote{An alternative approach is to introduce an additional stochastic variable that has delta-function localized correlation functions to remove these contact terms (see e.g.~\cite{Baumann:2010tm, McDonald:2006mx}).}

\subsection{Renormalized Halo Bias}
\label{sec:RHB}

We are now in a position to illustrate the basic idea of renormalized halo biasing. We will first do this with a specific example, the renormalization of $\delta^2$, leaving a complete treatment to the next section.  The results presented here have overlap with \cite{McDonald:2006mx, McDonald:2009dh, Chan:2012jx}.  Our general approach to renormalization agrees with these works, although it differs in detail.  One advantage of our approach is that it is systematic and can be carried out unambiguously order-by-order in perturbation theory.  
Moreover, we will clarify how the precise results depend on the choice of renormalization condition (and the renormalization scheme).  



\subsubsection{Example: Renormalization of $\delta^2$}
\label{sec:delta2}

Non-linear biasing contains terms with products of fields evaluated at the same point, e.g.~$\delta^n(\x)$, for $n>1$, in local Eulerian biasing~(\ref{equ:Fry}).
In field theory these terms are called {\it composite operators} and we will often use that terminology.
Composite operators can lead to additional divergences that cannot be removed by the renormalization of the dark matter density contrast---i.e.~by counterterms made out of (\ref{equ:EFTCT}).  
The simplest composite operator which appears in~(\ref{equ:Fry}) is~$\delta^2$. In this section, we will investigate the renormalization of this object. In fact, as we show in Appendix \ref{ap:HOO}, this will be the essential building block for the one-loop renormalization of the other composite operators appearing in the local Eulerian biasing model, namely $\delta^n$, for $n > 2$. 

\vskip 4pt
The composite operator  $\delta^2$ can be made finite by defining a new {\it renormalized operator} $[\delta^2]$ as
\beq
[\delta^2](\x, \tau) = \delta^2(\x, \tau) + \sum_{\widetilde{\cal O}}Z^{(\delta^2)}_{\widetilde{\cal O}}\,{\widetilde{\cal O}}(\x, \tau)\ ,\label{eq:delta2R}
\eeq
where the operators $\widetilde{\cal O}$ are counterterms introduced to absorb the UV divergences that arise in correlation functions of $\delta^2$. 
We need to distinguish the divergences which arise from non-linearities in the ``external legs''---i.e~in the dark matter contrast $\delta_{\q_i}$---and which are removed by the counterterms $\tilde\delta$ in the EFT-of-LSS, from the divergences which arise from contractions within the composite operator~$\delta^2$ and are renormalized by the counterterms in~(\ref{eq:delta2R}). 
To single out the internal divergences coming from the operator $\delta^2$, we replace the external legs $\delta_{\q_i}$ by their linear approximations $\delta^{(1)}_{\q_i}$ and impose the {\it renormalization conditions}
\beq
\vev{[\delta^2]_\q\,\delta_{\q_1}^{(1)}\cdots\,\delta_{\q_m}^{(1)}}=\vev{(\delta^2)_\q\,\delta_{\q_1}^{(1)}\cdots\,\delta_{\q_m}^{(1)}}_{\rm tree}\qquad{\rm at }\quad q_i= 0 \ , \, \forall\, i \ ,\label{eq:RC1}
\eeq
where the subscript ``tree" indicates the tree-level result.
Eq.~(\ref{eq:RC1}) therefore implies that the counterterms are chosen to precisely cancel the loop divergences
\beq
\sum_{\widetilde{\cal O}} Z^{(\delta^2)}_{\widetilde{\cal O}}\vev{\widetilde{\cal O}_\q\,\delta_{\q_1}^{(1)}\cdots\,\delta_{\q_m}^{(1)}}_{\rm tree}=-\vev{(\delta^2)_\q\,\delta_{\q_1}^{(1)}\cdots\,\delta_{\q_m}^{(1)}}_{\rm loop}\ , \label{equ:counter}
\eeq
where the subscript ``loop'' refers to diagrams which contain one contraction between two different fields within the composite operator $\delta^2$.  We see that, at one loop, the counterterms only need to be evaluated at tree level. 
We will discuss the renormalization conditions in more detail in \S\ref{ssec:RC}. 
Here, we will simply illustrate them in an example.

\vskip 4pt
We construct the renormalized operator $[\delta^2]$ order-by-order in perturbation theory.
To facilitate the calculation, we introduce the following diagrammatic representation: 
\beq
\delta^2\quad=\quad \includegraphicsbox[scale=0.8]{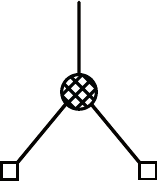} \quad \ ,
\eeq
where the ``blob'' indicates a convolution over the momenta.
\begin{itemize}
\item {\it m=0.---}The expectation value of $\delta^2$ is
\beq
\vev{(\delta^2)_\q}' =  \int_{0}^\Lambda\frac{\d p}{2\pi^2}\,p^2P(p) \equiv\sigma^2(\Lambda) \ ,
\eeq
where $\Lambda$ is a momentum cut-off introduced to regulate the loop.  Of course, this divergence (or cutoff-dependence) is removed by adding a constant counterterm
\beq
[\delta^2]=\delta^2-\sigma^2(\Lambda)\ .
\eeq
Subtracting this tadpole contribution ensures that $\vev{\delta_h}=0$ at the loop level. We note that the constant counterterm does not affect correlation functions with $m > 0$, as it only contributes to disconnected graphs.

\item {\it m=1.---}The one-loop contribution to $\vev{(\delta^2)_\q\,\delta_{\q_1}^{(1)}}$ is 
\beq
\vev{(\delta^2)_\q\,\delta_{\q_1}^{(1)}}'_{\rm loop} \quad = \quad 2\times\includegraphicsbox[scale=0.8]{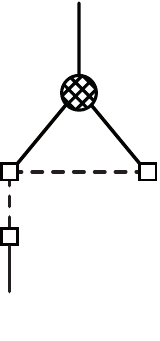}\quad =\quad \frac{68}{21}\sigma^2(\Lambda)\, P(q)\ ,\label{equ:1loop}
\eeq
We see that the dependence on the cut-off can be removed by a counterterm proportional to $\delta$, 
\beq
[\delta^2]\ =\ \delta^2 - \sigma^2(\Lambda)-\frac{68}{21}\sigma^2(\Lambda)\delta \quad \equiv \quad \includegraphicsbox[scale=0.8]{delta2}\quad \  + \ \ \quad\includegraphicsbox[scale=0.8]{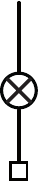}\quad \ \ ,\label{eq:ct1}
\eeq
where we have introduced a ``crossed circle'' to denote this new counterterm.

\item {\it m=2.---}Inserting (\ref{eq:ct1}) into $\vev{[\delta^2]_\q\,\delta_{\q_1}^{(1)}\delta_{\q_2}^{(1)}}$, we find
\begin{align}
\vev{[\delta^2]_\q\,\delta_{\q_1}^{(1)}\delta_{\q_2}^{(1)}}_{\rm loop}' \quad &= \quad  \includegraphicsbox[scale=0.8]{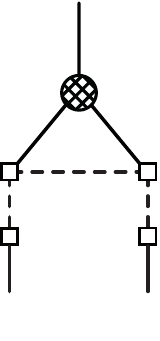}\quad+\quad 2\times\includegraphicsbox[scale=0.8]{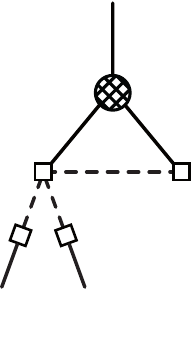} \quad + \quad\includegraphicsbox[scale=0.8]{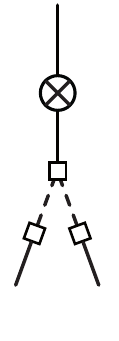}\quad .\label{eq:d22loop} \\
&= \quad \sigma^2(\Lambda)\left[\frac{5248}{735}+\frac{508}{2205}\frac{(\q_1\cdot\q_2)^2}{q_1^2q_2^2}+{\cal O}\Big(\frac{q_i^2}{\Lambda^2}\Big)\right]\,P(q_1)P(q_2)\ . \label{eq:div1}
\end{align}
While the first divergence can be absorb by a counterterm proportional to $\delta^2$, the second term cannot be removed by a counterterm which is local in $\delta$. Instead, the counterterm which removes this divergence is
\beq
\nabla_i\nabla_j\tilde\Phi_g\nabla^i\nabla^j\tilde\Phi_g\ ,\quad{\rm with}\quad\tilde\Phi_g\equiv\nabla^{-2}\delta =\frac{2}{3\H^2\Omega_m}\Phi_g\label{eq:rescaleg} \ ,
\eeq
where $\tilde\Phi_g$ is the {\it rescaled} gravitational potential. In order to keep the notation clean, we will drop the tilde and from now on denote the rescaled gravitational potential by $\Phi_g$. 
The renormalized operator $[\delta^2]$ then is
\beq
[\delta^2]=\delta^2-\sigma^2(\Lambda)\left[1+\frac{68}{21}\delta+\frac{2624}{735}\delta^2+\frac{254}{2205}(\nabla_i\nabla_j\Phi_g)^2\right]+\cdots\ ,\label{eq:ct2}
\eeq
where the ellipses refer to higher-derivative operators which are required to remove the subleading divergences of~(\ref{eq:div1}). 
Importantly, we see that the renormalization of the simplest operator in the local Eulerian biasing model {\it requires} the presence of an operator which is non-local in $\delta$. 



 \end{itemize}


\subsubsection{Comments}
\label{sec:Comments}

Before we move on to the more general treatment of renormalization, let us make a few comments:

\begin{itemize}
\item  The fact that renormalization requires non-local biasing is fundamentally different from the results of  \cite{Chan:2012jj}, which found that non-local bias is generated by time evolution.  In order to derive the latter result, one needs to  assume an equation for the time evolution of~$\delta_h$.  Instead, we are simply demanding that the predictions for correlation functions of halos are independent of how the theory is regulated.  Since the choice of regulator is arbitrary (and hence unphysical), our results follow only from the assumption that halos are physical.  Furthermore, since $\delta$ is itself a renormalized quantity, our counterterms are proportional to~$b_{n\geq 2}$ (see \S\ref{sec:RenBias}) and not $b_1$ as is found in \cite{Chan:2012jj}.    
\item The operator $\delta^2$ is a scalar under homogeneous boosts.  Therefore, $\delta^2$ can only be renormalized by operators which are also scalars under these transformations. 
From eq.~(\ref{equ:Phig}), we see that
the gravitational potential can only appear with (at least) two spatial derivatives acting on it, which is the case in~(\ref{eq:ct2}).  The gravitational potential itself only becomes an allowed operator if the symmetry is broken by the initial conditions~\cite{McDonald:2008sc} or if the equivalence principle is violated in the dynamics~\cite{Creminelli:2013nua, Kehagias:2013rpa}. 
\item At the order we have worked so far, the loop corrections have generated all operators which are scalars under boosts. In fact, this is quite general: loop corrections generate {\it every} operator consistent with the symmetries (here, the invariance under homogeneous boosts). Consequently, the bias relation should not only include the local operators~$\delta^n$, but every other operator allowed by these symmetries.
\item It is important to note that each diagram in~(\ref{eq:d22loop}), when evaluated separately, has divergences which cannot be removed by a scalar operator. In particular, we find divergences such as $q_i^{-2}$ or $q_i^{-2}(\q_1\cdot\q_2)$ which can only be removed by the operators  $ \Phi_g\delta$ and $\nabla_i \Phi_g\nabla^i\delta$. It is only when these three diagrams are summed that these undesirable divergences cancel.
\item It is convenient to rewrite the counterterms in (\ref{eq:ct2}) as
\beq
[\delta^2]=\delta^2-\sigma^2(\Lambda)\left[1+\frac{68}{21}\hskip 1pt\delta+\frac{8126}{2205}\hskip 1pt\delta^2+\frac{254}{2205}\hskip 1pt\G_2( \Phi_g)\right]+\cdots\ ,\label{eq:ct2x}
\eeq
where we defined 
\beq
\G_2( \Phi_g) \equiv (\nabla_i\nabla_j\Phi_g)^2 - (\nabla^2  \Phi_g)^2\ .\label{equ:galileon2}
\eeq
As we will explain in \S\ref{sec:Galileon}, and prove in Appendix \ref{app:Galileon}, the operator (\ref{equ:galileon2}) is part of a larger class of operators---the so-called {\it Galileon operators}~\cite{Nicolis:2008in,Chan:2012jj}---which are {\it not} renormalized at leading order in derivatives.  In other words, the divergences associated with these operators are only renormalized by derivative operators such as $\nabla^2\delta$.
\item We have shown explicit results at leading order in derivatives. Keeping the subleading terms, we find that higher-derivative operators are also required by renormalization
\beq
[\delta^2] \, \subset\,  -\, \tau^2(\Lambda) \left[\frac{1}{105} \frac{(\nabla \delta)^2}{\Lambda^2} +\frac{32}{245}\left(\frac{\nabla_i\nabla_j\Phi_g\nabla^i\nabla^j\delta}{\Lambda^2}-\frac{1}{3} \frac{\delta\nabla^2\delta}{\Lambda^2}\right)\right] \ ,\label{eq:ct2xx}
\eeq
where
\beq
\tau^2(\Lambda)\equiv\Lambda^2\int_0^\Lambda\frac{\d p}{2\pi^2}\, P(p)\ .
\eeq
Note that the operator $\nabla^2\delta$ does not appear at one-loop order. This is due to the fact that eq.~(\ref{equ:1loop}) is an exact result, and hence there is no divergence that scales as $q^2$. However, as we will explain in \S\ref{sec:Galileon}, the operator $\nabla^2\delta$ is generated by the one-loop renormalization of the Galileon operator $\G_2(\Phi_g)$.

\item The bias relation is an expansion both in small fluctuations and in derivatives. As we have explained, the expansion in small fluctuations makes sense only for renormalized operators. 
Similarly, the expansion in derivatives is only well-defined once the derivative terms have been appropriately renormalized. To be concrete, let us consider, at one-loop order, the operator $(\nabla \delta)^2$ and its correlation with $\delta^{(1)}$: 
\beq
\vev{(\nabla_i\delta \hskip 1pt \nabla^i\delta)_\q\,\delta^{(1)}_{\q'}}' = \left[\frac{82}{21}\gamma^2(\Lambda) +\frac{2}{3} q^2\sigma^2(\Lambda)\right]P(q)\ , \quad{\rm where}\quad\gamma^2(\Lambda)\equiv\int_0^\Lambda\frac{\d p}{2\pi^2}\, p^4P(p)\ .
\label{equ:Ddelta2}
\eeq
We see that the one-loop contribution from the {\it bare} operator $(\nabla\delta )^2$ does not vanish in the limit $q \to 0$. 
It is precisely 
these non-vanishing contributions that are removed when we define the renormalized operator $[(\nabla\delta )^2]$. 
Therefore, the derivative expansion is well-defined after the operators have been correctly renormalized.


\item The velocity potential starts appearing at cubic order through the following operator
\beq
\Gamma_3( \Phi_g,\Phi_v) \equiv \G_2(\Phi_g) - \G_2(\Phi_v)\ ,
\eeq
where $\Phi_v\equiv-\H^{-1}\nabla^{-2}\theta$ is the rescaled velocity potential and  the subscript ``3'' on $\Gamma_3$ was added to remind us that this is a cubic operator.
For lack of better term, we will call this operator the ``velocity tidal tensor". This operator is required to renormalize $\delta^2$ at $m=3$. 
 In \S\ref{sec:4pt}, we will see that the velocity potential leaves a distinct imprint in the halo trispectrum.

\end{itemize}

\section{One-Loop Renormalization of Halo Biasing}
\label{sec:Renorm2}

We now generalize the $\delta^2$-example of the previous section to a complete treatment of renormalized halo biasing.  In particular, we wish to extend our analysis to more general composite operators~${\cal O}$ (i.e.~other products of two or more fields at coincident points).   
The goal is to define a biasing model that is manifestly independent of the way the theory is regulated and write the bias expansion in terms of {renormalized operators} $[{\cal O}]$~\cite{collins1984renormalization, weinberg2005theV2}. 
By construction, correlation functions of $[{\cal O}]$ will be finite modulo divergences that are renormalized by the EFT-of-LSS and divergences that correspond to contact terms.
In \S\ref{sec:RenOp}, we describe the technical details associated with the construction of the basis of renormalized operators.  In \S\ref{sec:RenBias}, we present explicit results for the renormalized bias parameters at one-loop order. 

\subsection{Renormalized Operators}
\label{sec:RenOp}

The renormalized composite operators $[{\cal O}]$ are defined in terms of unrenormalized bare operators~${\widetilde {\cal O}}$ by~\cite{collins1984renormalization, weinberg2005theV2}
\beq
[{\cal O}] =  \sum_{\widetilde{\cal O}} {\cal Z}_{{\cal O},\widetilde{\cal O}}\,\widetilde{\cal O}\ ,
\label{equ:RenO}
\eeq 
where coefficients~${\cal Z}_{{\cal O},\widetilde{\cal O}}$ depend on the cutoff $\Lambda$, but the operators~$[{\cal O}]$ are independent of how the theory is regulated.
The counterterms on the right-hand side of (\ref{equ:RenO})  typically contain every term consistent with the symmetries.
We will first list the available terms, up to third order in perturbation theory and at lowest order in derivatives. As we will see, this analysis is complicated by the fact that some operators are related by the equations of motion and are therefore not independent. For related discussion, see~\cite{McDonald:2009dh, Chan:2012jj, Kehagias:2013rpa}.



\subsubsection{Symmetries and Counterterms}
\label{sec:counter}

In \S\ref{sec:fluideqs}, we have seen that the gradients of the gravitational potential $\Phi_g$ and the velocity potential $\Phi_v$ shift by a vector under homogeneous boosts. If the initial conditions are also invariant under such boosts, the allowed counterterms should only involve  operators which are scalars under these transformations. Consequently, the basic building block for constructing the renormalized theory is
\beq
\nabla_i\nabla_j\Phi\ ,\quad{\rm where}\quad \Phi= \Phi_g\ {\rm  or}\ \Phi_v\ . \label{eq:op1}
\eeq
Although we will mostly work at leading order in derivatives, the results of this section can be generalized straightforwardly to include higher-derivative terms. 
Rotational invariance implies that the indices in~(\ref{eq:op1}) need to be contracted. 
At first order, the bias relation can therefore depend on 
\beq
\nabla^2\Phi_g\ , \ \nabla^2\Phi_v\ , \label{equ:Op}
\eeq
while at second order, it may contain the following terms
\beq
(\nabla^2\Phi_g)^2\ ,\ (\nabla^2\Phi_v)^2\ ,\ \nabla^2\Phi_g\nabla^2\Phi_v\ ,\ (\nabla_i \nabla_j\Phi_g)^2\ ,\ (\nabla_i \nabla_j\Phi_v)^2\ , \ (\nabla_i \nabla_j\Phi_g)(\nabla^i \nabla^j\Phi_v)\ .\label{equ:Op2}
\eeq
At a given order in perturbation theory, the operators in (\ref{equ:Op}) and (\ref{equ:Op2}) are not all independent, but are related by the dark matter equations of motion:

\begin{itemize}
\item {\it 1st order.}---At linear order, the (rescaled) gravitational and velocity potentials are equal
\beq
\Phi_g^{(1)}=\Phi_v^{(1)}\ .
\eeq
Hence, there is just one independent operator in (\ref{equ:Op}), which we choose to be $\delta = \nabla^2 \Phi_g$.
 
 \item {\it 2nd order.}---Since $\Phi_g^{(1)}=\Phi_v^{(1)}$, we do not distinguish between the gravitational potential and the velocity potential in the list of quadratic operators in~(\ref{equ:Op2}), which therefore only contains two independent operators: $\delta^2 = (\nabla^2 \Phi_g)^2$  and $\G_2(\Phi_g) = (\nabla_i \nabla_j \Phi_g)^2 - (\nabla^2 \Phi_g)^2$.
Moreover, it is easy to show from (\ref{equ:F2}) and (\ref{equ:G2}) that, at second order, the difference between the density contrast and the velocity divergence is the Galileon operator~(\ref{equ:galileon2}),
\beq
\nabla^2\Phi_g^{(2)} - \nabla^2\Phi_v^{(2)} = -\frac{2}{7}\hskip 1pt \G_2(\Phi^{(1)}_g )\ .
\eeq
Hence, although a priori eight operators are consistent with the symmetries, only three are independent after using the second-order equations of motion. 
We chose these independent operators to be 
\beq
\delta\ ,\ \delta^2\ , \ \G_2(\Phi_g)\ .
\eeq

\item {\it 3rd order.}---At cubic order, the velocity potential appears for the first time as an independent degree of freedom. Indeed, a set of independent operator at this order is \cite{Chan:2012jj}
\beq
\delta\ ,\ \delta^2\ ,\ \delta^3\ ,\ \G_2(\Phi_g)\ ,\ \G_2(\Phi_g)\delta\ ,\ \G_3(\Phi_g)\ , \ \Gamma_3(\Phi_g, \Phi_v) \equiv \G_2(\Phi_g)-\G_2(\Phi_v) \ , \label{equ:idptOp}
\eeq 
where $\G_3(\Phi_g)$ is the third-order Galileon operator 
\beq
\G_3( \Phi_g)\equiv-\frac{1}{2}\left[2\nabla_i\nabla_j  \Phi_g \nabla^j\nabla_k  \Phi_g\nabla^k\nabla^i  \Phi_g + (\nabla^2  \Phi_g)^3-3(\nabla_{i}\nabla_{j} \Phi_g)^2\nabla^2  \Phi_g\right]\ , \label{equ:galileon3}
\eeq
and $\Gamma_3$ contains the velocity tidal tensor $\G_2(\Phi_v)$.

\end{itemize}
Of course, this analysis can, in principle, be carried out to any order in perturbation theory. 
However, the set of independent operators listed in (\ref{equ:idptOp}) will be sufficient for most of this paper (but see Appendix~\ref{app:Bispectrum}).  We will formulate renormalization in terms of this basis of operators.

\subsubsection{Renormalization Conditions}
\label{ssec:RC}

\begin{figure}[h!]
   \centering
       \includegraphics[scale =0.9]{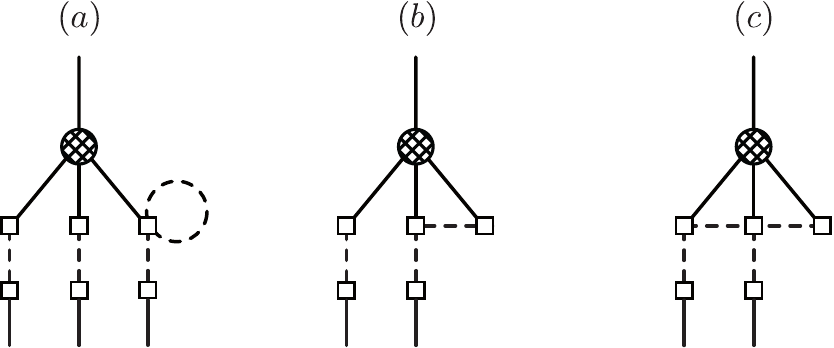}
       \caption{Diagram $(a)$ is not a (partially) 1PI diagram, as it does not contain contractions between different fields within the composite operator. On the other hand, diagrams $(b)$ and $(c)$ are examples of a partial 1PI graph and a full 1PI graph, respectively. }
       \label{fig:1PI}
\end{figure}

We are interested in correlation functions of the form $\vev{{\cal O}_\q\,\delta_{\q_1}\cdots\delta_{\q_m}}$.
These correlation functions contain two types of divergences. Those associated with non-linearities in the external $\delta$'s (which are renormalized by the EFT-of-LSS) and those {\it within} the operator~${\cal O}$.
Since we will be interested in the latter, we may replace the external legs~$\delta_{\q_i}$ by their linear approximations $\delta^{(1)}_{\q_i}$ and restrict to  (partially) {\it one-particle irreducible} (($p$)1PI) diagrams (see fig.~\ref{fig:1PI}).

In order to fix the finite contributions in the counterterms in (\ref{equ:RenO}), we impose renormalization conditions. We will require that the counterterms exactly cancel the divergences on large scales, i.e.~in the limit where the external momenta  vanish. In other words, we define the renormalized operators $[\cal O]$ through\hskip 1pt\footnote{These renormalization conditions are enough to remove divergences at {leading order} in derivatives. If the correlation functions
$\vev{[{\cal O}]_\q\, \delta_{\q_1}^{(1)}\cdots\delta_{\q_m}^{(1)}}$
contain divergences proportional to positive powers of the external momenta, additional renormalization conditions need to be imposed on derivatives of the (amputated) correlation functions with respect to these momenta.}
\beq
\vev{[{\cal O}]_\q\, \delta_{\q_1}^{(1)}\cdots\delta_{\q_m}^{(1)}}_{\rm {\it (p)}1PI}=0\qquad{\rm at }\quad q_i= 0 \ , \, \forall\, i  \ ,\label{eq:RCgeneral}
\eeq
where the subscript ``({\it p})1PI'' denotes (partially) 1PI diagrams.  In Appendix~\ref{sec:mixing}, we discuss the consequences of imposing the renormalization conditions at finite momentum, $q_i =  \mu$.
In particular, we show that the basis of renormalized operators that is orthogonal at a scale $\mu$ will mix and in general won't be an orthogonal basis at another scale $\mu'$.
Related to this is the fact that the renormalized bias parameters (see \S\ref{sec:RenBias}) depend on the renormalization scale $\mu$.

One may be concerned that the renormalization conditions~(\ref{eq:RCgeneral}) are not enough to ensure that every UV divergence which can appear in $\vev{{\cal O}_\q\,\delta_{\q_1}\cdots\delta_{\q_m}}$ has been removed. Indeed, the physical object we wish to compute is not the 1PI part of the correlation of $\cal O$ with the linear dark matter contrast, but the connected part of the correlation function $\vev{{\cal O}_\q\,\delta_{\q_1}\cdots\delta_{\q_m}}$.
However, since at any given order in perturbation theory the non-linear dark matter contrast $\delta$ can be written as a product of  linear dark matter contrasts $\delta^{(1)}$, these (partially) 1PI diagrams are the building blocks of correlation functions with the non-linear dark matter density contrast. Any remaining loops are either renormalized by counterterms in the EFT-of-LSS or are finite up to contact terms.

\subsubsection{Non-Renormalization of Galileon Operators}
\label{sec:Galileon}

In eqs.~(\ref{equ:galileon2}) and (\ref{equ:galileon3}), we defined the Galileon operators $\G_2$ and $\G_3$. These definitions can be extended to $n$-th order Galileon operators, whose precise expressions can be found in Appendix~\ref{app:Galileon}.
Importantly, every Galileon operator $\G_n(\Phi)$ can be written as the  second derivative of another operator $T_{ij}^{(n-1)}$,
\beq
\G_n(\Phi)\, =\, \nabla^i\nabla^j\,\big(\Phi\,T_{ij}^{(n-1)}\big)\ ,
\eeq
As a result, the correlation of $\G_n$ with $(\delta^{(1)})^m$ scales as $q^2$, provided that $\vev{[\Phi\,T_{ij}^{(n-1)}]_\q\, \delta_{\q_1}^{(1)}\cdots \delta_{\q_m}^{(1)} }$ isn't singular when any of the external momenta (or partial sums of the external momenta) go to zero. As we explain in Appendix~\ref{app:Galileon}, the absence of such poles is guaranteed by symmetry considerations. 
Consequently, loops arising from Galileon operators are only renormalized by higher-derivative operators
\beq
[\G_n(\Phi)]\, =\, \G_n(\Phi)\, +\, {\cal O}\Big(\frac{\nabla^2}{\Lambda^2}\Big)\ .
\eeq
At leading order in derivatives, but to all orders in loops, the Galileon operators $\G_n$ are not renormalized.

\subsection{Renormalized Bias Parameters}
\label{sec:RenBias}

In terms of the bare operators, the halo density contrast is
\beq
\delta_h = \sum_{\cal O} b_{\cal O}^{(0)} {\cal O} \ .
\eeq
If the expansion contains all operators consistent with the symmetries, then it can be re-written in the basis of renormalized operators
\beq
\delta_h = \sum_{\cal O} b_{\cal O}^{(R)} [{\cal O}] \ ,
\eeq
where $b_{\cal O}^{(R)}$ are the renormalized bias parameters.
We can gain intuition for the form of the renomalized bias parameters from the results of \S\ref{sec:RHB}.
Consider the bare expansion at quadratic order, $\delta_h = b_{0}^{(0)}  + b_1^{(0)} \delta +  \frac{1}{2}b_2^{(0)} \delta^2 + b_{\G_2}^{(0)} \G_2+ \cdots$.  To write this in terms of $[\delta^2]$, we add and subtract the counterterms in eq.~(\ref{eq:ct2x}). 
This shifts $b_0^{(0)}$, $b_1^{(0)}$, $b_2^{(0)}$ and $b_{\G_2}^{(0)}$ by a term proportional to $b_2^{(0)} \sigma^2(\Lambda)$.  The simple lesson is that the renormalized bias coefficients corresponding to operators~${\cal O}_I$ are shifted from their bare values by contributions proportional to the bare bias parameters of any operators ${\cal O}_J$ that require ${\cal O}_I$ as counterterms. 
Since $\delta$ is finite by the EFT-of-LSS, the bias parameter $b_1^{(0)}$ will not appear in any renormalized bias coefficient (other than $b_1^{(R)}$), but $b_1^{(R)} \neq b_1^{(0)}$ because $\delta$ is needed as a counterterm for many renormalized operators.
Similarly, since the Galileon operators aren't renormalized at leading order in derivatives the renormalized bias coefficients at this order do {\it not} depend on the bare parameters of the Galileon operators.

\vskip 4pt
More formally, the relationship between the renormalized bias parameters $b^{(R)}_{\cal O}$ and the bare bias parameters~$b^{(0)}_{\cal O}$~is
\beq
b_{\cal O}^{(R)}\equiv\sum_{\widetilde{\cal O}}{\cal Z}_{\widetilde{\cal O},{\cal O}}^{-1}\,b_{\widetilde{\cal O}}^{(0)}\ ,\label{eq:Rbias}
\eeq
where ${\cal Z}_{\widetilde{\cal O},{\cal O}}^{-1}$ is the inverse of the matrix that appears in (\ref{equ:RenO}).
To determined  $b_1^{(R)}$ to order $\sigma^2$, we need to renormalized all cubic operators (up to $m=1$), except the Galileon operators $\G_2$ and $\G_3$ which aren't renormalized at leading order in derivatives.
This calculation is performed in Appendix~\ref{ap:HOO}.
We find
\beq
b_1^{(R)}\, =\, b_1^{(0)}\, +\, \sigma^2(\Lambda) \left[\frac{34}{21}b_2^{(0)}+\frac{1}{2}b_3^{(0)} -\frac{4}{3}b_{\G_2\delta}^{(0)}\right]  \ .\label{eq:b1renorm}
\eeq
To consistently renormalize the quadratic bias parameters $b_2^{(R)}$ and $b_{\G_2}^{(R)}$, quartic operators need to be taken into account (and renormalized up to $m=2$).
After a lengthy computation, we get
\beq
b_2^{(R)} = b_2^{(0)} + \sigma^2(\Lambda)\, \bigg[\,\frac{8126}{2205}b_2^{(0)}+\frac{68}{21} b_{3}^{(0)} -\frac{752}{105} b_{\G_2\delta}^{(0)} +\frac{1}{2} b_4^{(0)} - \frac{16}{3}b^{(0)}_{\G_2\delta^2}-\frac{128}{105}b_{\Gamma_3\delta}^{(0)}+\frac{64}{15} b_{{(\G_2)^2}}^{(0)}\bigg] \ ,\label{eq:b2renorm}
\eeq
and 
\beq
b_{\G_2}^{(R)} =   b_{\G_2}^{(0)} + \sigma^2(\Lambda)\, \left[\, \boxed{\frac{127}{2205}b_2^{(0)}}\, +\frac{116}{105}b_{\G_2\delta}^{(0)}+\frac{1}{2}b_{\G_3\delta}^{(0)}+b^{(0)}_{{\G_2\delta^2}} +\frac{8}{35}b_{{\Gamma_3\delta}}^{(0)}+\frac{8}{15}b_{{(\G_2)^2}}^{(0)}\right] \ .\label{eq:beta2renorm}
\eeq
Notice that, at one-loop order, the operators $\delta^3$ and $\delta^4$ do not generate a divergence proportional to the Galileon operator~$\G_2$, so the bare bias parameters $b_3^{{(0)}}$ and  $b_4^{{(0)}}$ are absent from~$b_{{\G_2}}^{(R)}$.  However,  the operator $\delta^2$ does contribute to the running of $b_{\G_2}^{(0)}$.
 This means that, even if the non-local bias parameter is set to zero at some scale $\Lambda$---as in the case of the local Eulerian biasing model---this will no longer be true at some other scale $\Lambda^\prime$. Consequently, the local Eulerian biasing model is not a consistent model beyond the tree-level approximation.  Finally, we note that the dependence of $b_{{\G_2}}^{(R)}$ on the bare bias parameters is quite different from the result of \cite{Chan:2012jj}, where the time-evolved Eulerian bias parameter, $b_{{\G_2}}$, was related to the linear Lagrangian bias parameter, $b_1^{(0)}$, at some earlier time.  Since $\delta$ is a renormalized operator in the EFT-of-LSS, $b_{1}^{(0)} $ does not appear in $b_{\G_2}^{(R)}$ (or any other renormalized bias parameter).

\vskip 4pt

\section{Halo Statistics}
\label{sec:HaloStatistics}

Self-consistent renormalization has forced us to consider a biasing model of the form
\beq
\delta_h = \sum_{\cal O} b_{\cal O}^{(R)} [{\cal O}]\ , \label{equ:BiasEFT}
\eeq
where the right-hand side is a double expansion in small fluctuations ($\Phi_g$ and $\Phi_v$) and spatial derivatives~($\nabla_i$). After renormalization, the higher-derivative terms are suppressed by $\Lambda_\star$, the physical scale of non-locality in halo formation. For dark matter halos, we expect\footnote{Recall that dark matter particles have travelled less than the non-linear distance $\knl^{-1}$ over the history of the universe~\cite{Baumann:2010tm}.} $\Lambda_\star^{-1} \le \knl^{-1}$, but ultimately $\Lambda_\star$ should be determined from N-body simulations or observational data.
The renormalized biasing model (\ref{equ:BiasEFT}) should be viewed as an effective theory valid on scales larger than~$\Lambda_\star^{-1}$.
As is typically for effective theories, only a finite number of terms need to be retained in  (\ref{equ:BiasEFT}) in order to describe halo statistics to a finite accuracy.
In \S\ref{sec:PC}, we perform a simple power counting to estimate the relative sizes of the renormalized operators. In \S\ref{sec:2point} and \S\ref{sec:HPF}, we describe how the lowest-order bias parameters can be measured by fitting the predictions of~(\ref{equ:BiasEFT}) to a variety of halo correlation functions (either in N-body simulations or in observational data).
Readers who are less interested in the technical details may find summaries of results in \S\ref{sec:summary1} and~\S\ref{sec:summary}.

\subsection{Power Counting}
\label{sec:PC}

We begin with an estimate of the relative sizes of the renormalized operators. 
 In the absence of derivatives of $\delta$, we may simply count powers of the linear dark matter density contrast $\delta^{(1)}$. 
The relative contribution from higher-derivative terms depends on the initial statistics.\footnote{In the EFT-of-LSS, it has been shown that higher-derivative terms become important as one approaches the non-linear scale~\cite{Carrasco:2012cv}.}  In \cite{Pajer:2013jj, Carrasco:2013mua}, the contribution from derivatives were estimated using a power law ansatz for the dark matter spectrum
\beq
\Delta^2_\delta(q) = \frac{q^3}{2\pi^2} P(q) \sim \left(\frac{q}{k_{\mathsmaller{\rm NL}}}\right)^{3+n}\ ,
\eeq
where $n$ is scale-dependent, varying from $n\simeq - 2.1$~\cite{Carrasco:2013mua} near the non-linear scale (at $z=0$) to $n \approx 1$ on large scales.  Since the higher-order biasing terms are mostly relevant near the non-linear scale, we will use $n = - 2$ for our estimates.
In that case, $\frac{1}{2}(3+n) = \frac{1}{2}$ and each power of $\delta$ roughly adds a power of~$q^{1/2}$.   
To be conservative about the relevance of higher-derivative operators, we will use $\Lambda_\star \sim \knl$. In reality, we expect $\Lambda_\star > \knl$ and higher-derivative contributions will be more suppressed than what we estimate here.
Adopting this power counting for suitably renormalized operators, we get
 \beq
\left\{ [{\cal O}_{(\alpha,\beta)}]\right\} \equiv \left\{ \frac{[\nabla^{2\alpha} (\nabla^2 \Phi_g)^\beta]}{\Lambda_\star^{2\alpha} } \right\} \, \sim\, \frac{q^{2\alpha}}{\Lambda_\star^{2\alpha}} \Delta_\delta^\beta \, \sim \, \frac{q^{2\alpha}}{\Lambda_\star^{2\alpha}} \left( \frac{q}{\knl}\right)^{\frac{1}{2}(3+n) \beta} \, \sim\, \left( \frac{q}{\knl} \right)^{2\alpha + \frac{1}{2}\beta} \ .
 \eeq
 where in the last equality we have used $\Lambda_\star \sim \knl$ and $\frac{1}{2}(3+n)  \sim \frac{1}{2}$.
At leading order in derivatives, we therefore have
 \begin{align}
 q^{1/2}: \qquad {\cal O}_{(0,1)} &= \left\{\, \delta \, \right\} \ , \label{equ:O1} \\
  q^{1}: \qquad  {\cal O}_{(0,2)} &= \left\{\, \delta^2  \ , \ {\cal G}_2  \, \right\} \ , \\
  q^{3/2}: \qquad  {\cal O}_{(0,3)} &= \left\{\, \delta^3   \ , \ {\cal G}_3  \ , \   \G_2 \delta\ ,\ \Gamma_3  \, \right\} \ .
  \end{align}
Shown here is the leading $q$-scaling in the long-wavelength limit.  When these operators are inserted into correlation functions they may be further suppressed due to the fact that certain correlations vanish for Gaussian initial conditions.
 Higher-derivative operators enter at higher order in the $q/\knl$ expansion:
 \begin{align} 
\hspace{-1.47cm} q^{5/2}: \qquad  {\cal O}_{(1,1)} &= \left\{\, \nabla^2\delta  \ , \ \cdots \right\} \ . \label{equ:O4}
 \end{align}
Of course, when the operators in (\ref{equ:O1})--(\ref{equ:O4}) are inserted in the bias expansion (\ref{equ:BiasEFT}) their relative contributions will depend on the relative sizes of the renormalized bias parameters $b_{\cal O}^{(R)}$.  A large hierarchy between certain bias parameters can affect the estimates that we have performed here.
Hence, our estimates should only be viewed as qualitative guidelines, and a more detailed treatment (analogous to~\cite{Carrasco:2013mua}) is clearly required for comparisons with real data.

\subsection{Two-Point Statistics}
\label{sec:2point}

In this section, we compute the halo-matter power spectrum, $P_{hm}(q)\equiv \vev{(\delta_h)_{\q}\hskip 1pt\delta_{\q'}}'$, and the halo-halo power spectrum, $P_{hh}(q)\equiv \vev{(\delta_h)_{\q}\hskip 1pt(\delta_h)_{\q'}}'$,
at one loop (i.e.~to fourth order in $\delta^{(1)}$).  We will present results only at leading order in derivatives, but it will be clear how higher-derivative operators would be included. The (renormalized) operators which give non-vanishing one-loop contributions to $P_{hm}$ and $P_{hh}$ then are: 
\beq
\delta \quad {\rm and} \quad  [{\cal O}] \equiv \{\, [\delta^2]\ ,\ [\G_2]\ ,\ [\Gamma_3]\, \}\ . \label{equ:OP}
\eeq 
The non-linear dark matter power spectrum, $P_{mm}(q)\equiv\vev{\delta_\q\hskip 1pt\delta_{\q'}}'$, is renormalized in the EFT-of-LSS, cf.~eq.~(\ref{equ:PNL}). 
The contributions from the remaining
 operators~$[{\cal O}]$ will be discussed in this section.

\subsubsection{Halo-Matter} 
\label{sec:hm}

Diagrammatically, the correlation between $[{\cal O}]$ and $\delta$ is
\begin{align}
\langle [{\cal O}]_{\q}\hskip 1pt \delta_{\q'} \rangle' \quad &\ =\quad  \ \ \includegraphicsbox[scale=0.9]{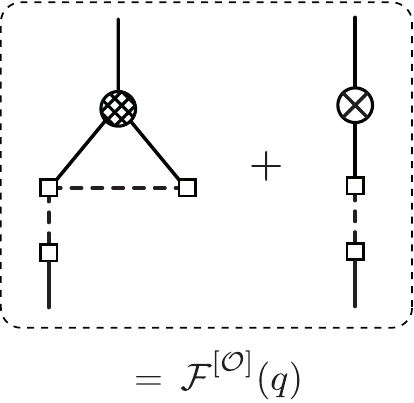}\qquad+\qquad \includegraphicsbox[scale=0.9]{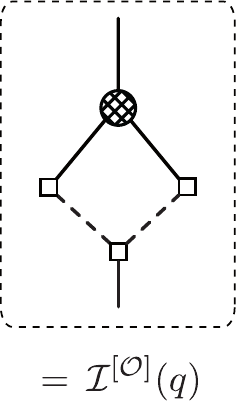}\qquad .
\end{align}
We have separated $\langle [{\cal O}]_{\q}\hskip 1pt \delta_{\q'} \rangle'$ into ${\cal F}$-terms, which contain a contraction between the two legs of the composite operators, and ${\cal I}$-terms, which only contain contractions with the external leg.
The halo-matter power spectrum (\ref{equ:Phm}) can then be written as~\cite{McDonald:2009dh}
\beq
P_{hm}(q)=  b_1^{(R)} \hskip 1pt P_{mm}(q) + \sum_{\cal O} b_{{\cal O}}^{(R)} \Big({\cal F}^{[{\cal O}]}(q) + {\cal I}^{[{\cal O}]}(q) \Big) \ , \label{equ:Phm}
\eeq
where we have defined $b_{\delta^2}^{(R)} \equiv \frac{1}{2} b_{2}^{(R)}$.

\begin{itemize}
\item {${\cal F}${\it-terms.}}---The functions ${\cal F}^{[{\cal O]}}$  
are 
\begin{align}
{\cal F}^{[\delta^2]}(q)&\equiv 0 \ , \\
{\cal F}^{[\G_2]}(q)&\equiv 4P(q)\int_\p \sigma_{\p,\q-\p}^2 \hskip 2pt F_2(\q,-\p)\, P(p) \ ,\label{equ:FG2} \\
{\cal F}^{[\Gamma_3]}(q)&\equiv  - \frac{8}{7}P(q)\int_\p \sigma_{\p,\q-\p}^2  \hskip 1pt \sigma_{\p,\q}^2 \hskip 1pt P(p) \ ,\label{equ:FGamma3}
\end{align}
where $\sigma^2_{\p,\q}\equiv (\p\cdot\q/pq)^2-1$.  The two non-zero ${\cal F}$-terms are proportional to each other
\beq
{\cal F}^{[\G_2]}(q) = \frac{5}{2}{\cal F}^{[\Gamma_3]}(q) \equiv {\cal F}(q) \equiv f(q) P(q) \ .
\eeq
The two operators $\G_2$ and $\Gamma_3$ therefore contribute degenerate ${\cal F}$-terms to the halo-matter power spectrum.
Since the ${\cal F}$-terms are proportional to the linear power spectrum they be interpreted as a scale-dependent contribution to the linear bias
\beq
b_L^{(R)}(q) \equiv b_1^{(R)} + \left(b_{\G_2}^{(R)} + \frac{2}{5}b_{\Gamma_3}^{(R)} \right) f(q) + \cdots\ .
\eeq

In the case of a scaling universe with $P(q)\propto q^n$, the finite part of ${\cal F}$ can be computed analytically using dimensional regularization \cite{Pajer:2013jj}  
\beq
f(q)= c(n)\, \Delta_\delta^2(q)\ , 
\eeq
where $c(n)$ is a coefficient which depends on the scaling of the power spectrum.
 In our universe, the integrals in (\ref{equ:FG2}) and (\ref{equ:FGamma3}) are convergent as the cut-off $\Lambda$ is taken to infinity. Of course, by sending $\Lambda$ to infinity one is including contributions from scales which are outside the regime of validity of the effective description. However, it is easy to see that these finite errors can be absorbed into the bias parameters of higher-derivative operators. After renormalization, we expect the higher-derivative operators to be suppressed by a (momentum) scale that is larger than the scale of variation of the functions in (\ref{equ:FG2}) and~(\ref{equ:FGamma3}).

\item {${\cal I}${\it-terms.}}---The functions ${\cal I}^{[{\cal O}]}$ in~(\ref{equ:Phm}) are
\begin{align}
{\cal I}^{[\delta^2]}(q)&\equiv 2\int_\p F_2(\q-\p,\p)\, P(p)P(|\q-\p|)\ ,\label{equ:Ihmd2}\\
{\cal I}^{[\G_2]}(q)&\equiv 2\int_\p \sigma_{\p,\q-\p}^2  \hskip 2pt  F_2(\q-\p,\p)\, P(p)P(|\q-\p|)\ ,\label{equ:IhmG2} \\
{\cal I}^{[\Gamma_3]}(q)&\equiv 0 \ .
\end{align}
Although the integrals in (\ref{equ:Ihmd2}) and (\ref{equ:IhmG2}) receive contributions from all scales, they are  {finite} up to contact terms.  More precisely, these integrals can be written as the sum of a finite term and a cut-off dependent term,
\beq
{\cal I}^{[{\cal O}]}(q) = I^{[{\cal O}]}\left(\frac{q}{k_{\mathsmaller{\rm NL}}}\right)+J^{[{\cal O}]}\left(\Lambda,\frac{q^2}{\Lambda^2}\right)\ .
\eeq
Importantly, the functions $J^{[{\cal O}]}(q)$ are analytic in $q^2$, i.e.~they can be written as expansions in powers of $(q^2)^n$. In position space, these terms become derivatives of delta functions and therefore disappear when correlation functions are evaluated at separated points. The cut-off dependent parts are therefore {contact terms} which can be safely discarded and only the physical finite terms $I^{[{\cal O}]}(q)$ are kept. 
In a scaling universe, the finite parts can again be computed in dimensional regularization 
\beq
{\cal I}^{[{\cal O}]}(q)=d^{[{\cal O}]}(n)\, \Delta_\delta^{2}(q)\, P(q)\ , 
\eeq
where $d^{[{\cal O}]}(n)$ are coefficients which depend on the scaling $n$ of the power spectrum.
In the real universe,  the two functions ${\cal I}^{[\delta^2]}$ and ${\cal I}^{[\G_2]}$ are approximately proportional to each other,
 \beq
 {\cal I}^{[\delta^2]}(q) \approx -\frac{5}{4}\, {\cal I}^{[\G_2]}(q) \equiv {\cal I}(q)\ ,
 \eeq
as illustrated in fig.~\ref{fig:Phm}. The operators $\delta^2$ and $\G_2$ therefore contribute degenerate ${\cal I}$-terms.

\begin{figure}[h!]
   \centering
       \includegraphics[width=0.75\textwidth]{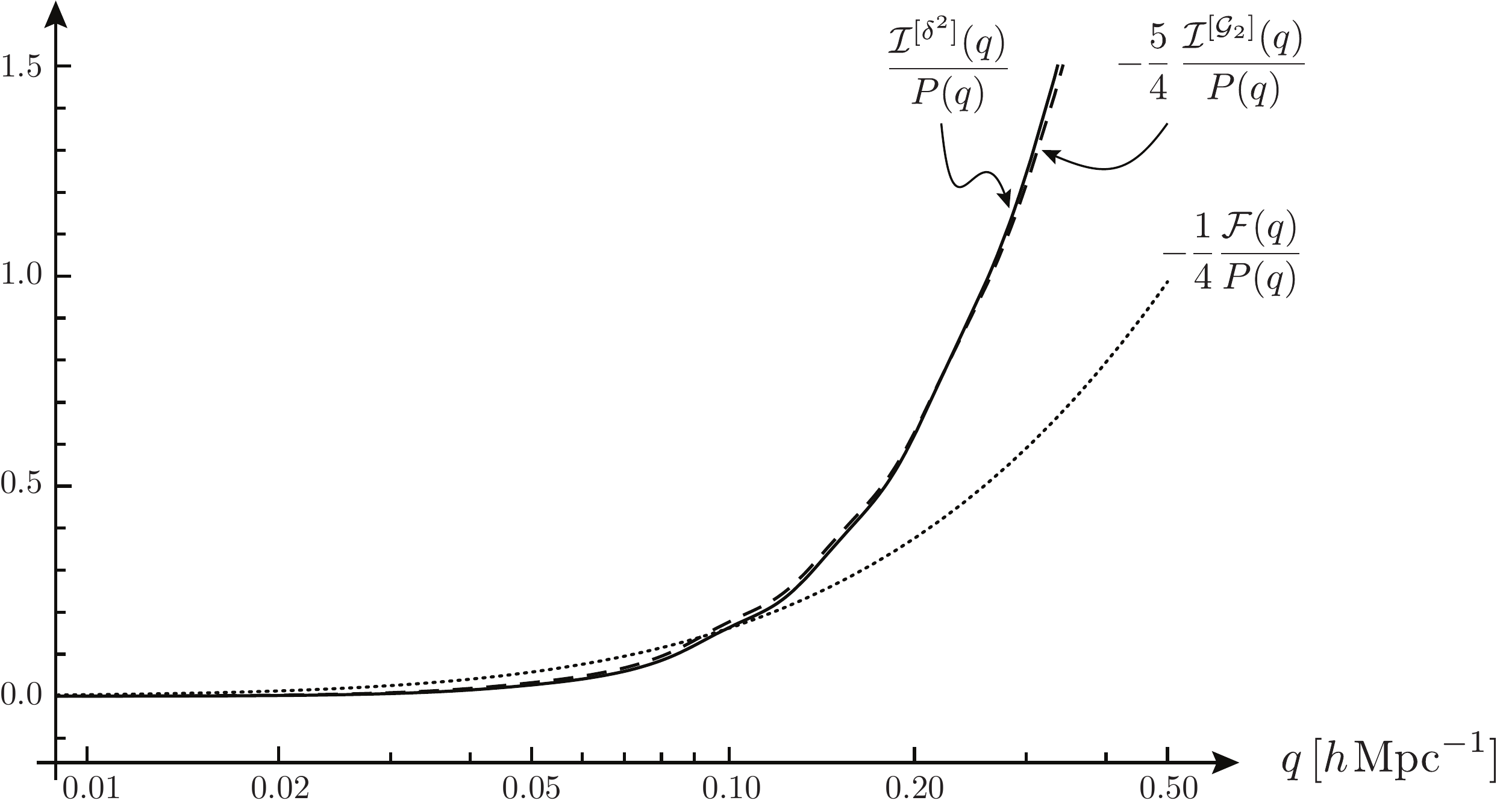}
    \caption{Numerical results for the ${\cal F}$- and ${\cal I}$-terms of halo-matter power spectrum, $P_{hm}(q)$, in the real universe. To a good approximation the ${\cal I}$-terms are proportional to each other.}
  \label{fig:Phm}
\end{figure}

\end{itemize}

\subsubsection{Halo-Halo} 

\begin{figure}[h!]
   \centering
       \includegraphics[width=0.75\textwidth]{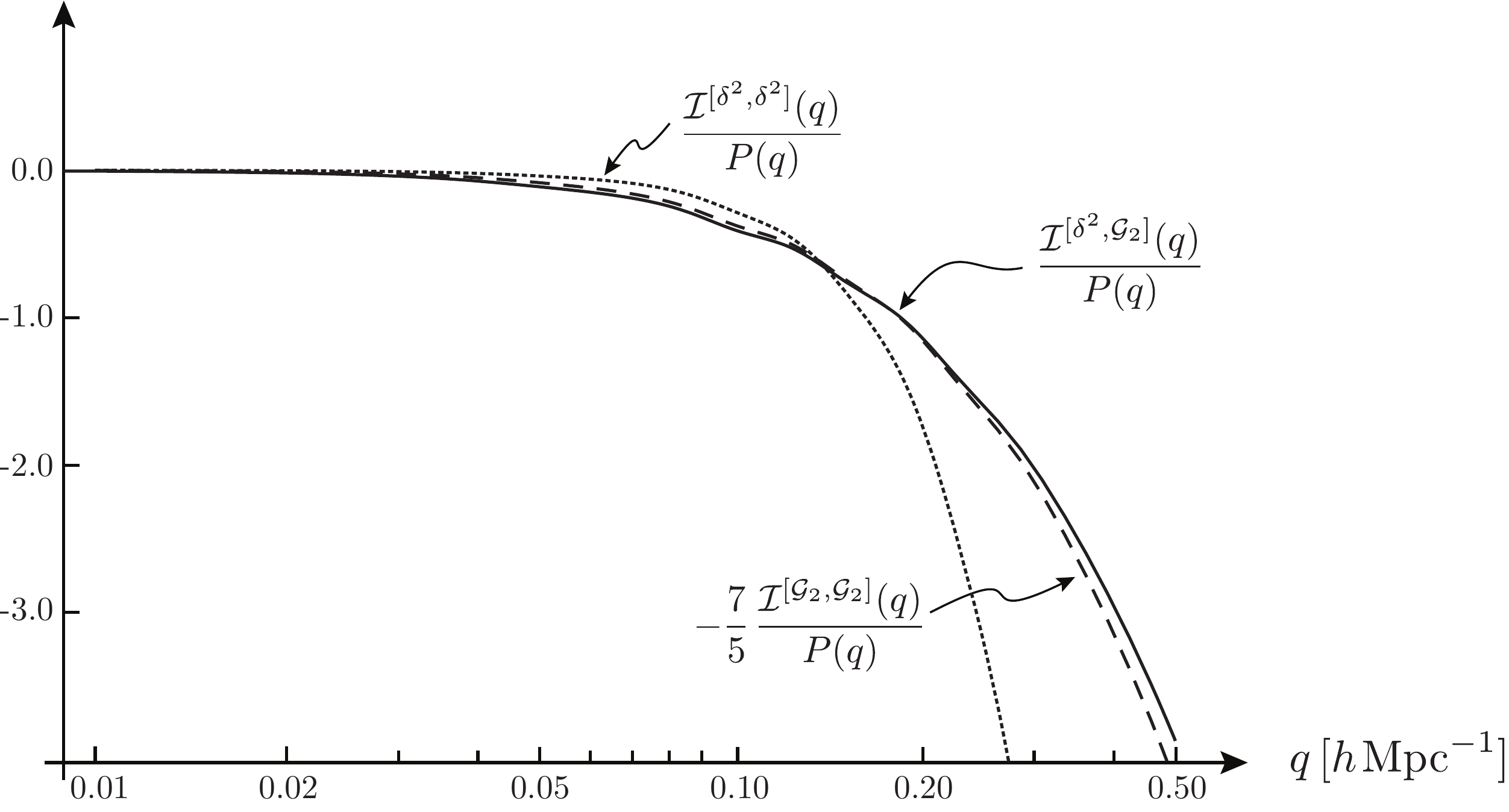}
   \caption{Numerical results for the ${\cal I}$-terms of halo-halo power spectrum, $P_{hh}(q)$, in the real universe.}
  \label{fig:Phh}
\end{figure}

A similar one-loop calculation for the halo-halo power spectrum gives~\cite{McDonald:2009dh} 
\beq
P_{hh}(q) = b_1^{(R)} \left[b_1^{(R)}P_{mm}(q)+2\sum_{\cal O}b_{\cal O}^{(R)}\left({\cal F}^{[{\cal O}]}(q)+{\cal I}^{[{\cal O}]}(q)\right)\right] + \sum_{{\cal O}, {\cal O}'}b_{{\cal O}}^{(R)} b_{{\cal O}'}^{(R)}\, {\cal I}^{[{\cal O},{\cal O}']}(q)\ , \label{equ:Phh2}
\eeq
where the functions ${\cal I}^{[{\cal O},{\cal O}']}(q)\equiv \vev{[{\cal O}]_{\q}\hskip 1pt[{\cal O}']_{\q'}}' $ are
\begin{align}
{\cal I}^{[\delta^2, \delta^2]}(q)&= 2\int_{\p}P(p)P(|\q-\p|)\ ,\\
{\cal I}^{[\G_2,\G_2]}(q)& =2\int_{\p} (\sigma_{\p,\q-\p}^2)^2 \hskip 1pt P(p)P(|\q-\p|)\ ,\\
{\cal I}^{[\delta^2,\G_2]}(q)& =2\int_{\p} \sigma_{\p,\q-\p}^2 \hskip 2pt P(p)P(|\q-\p|)\ , \\
{\cal I}^{[\Gamma_3,{\cal O}']}(q)& = 0 \ .
\end{align}
As before, these integrals are finite up to contact terms.  In a scaling universe, these finite parts simply are
\beq
{\cal I}^{[{\cal O},{\cal O}']}(q)=d^{[{\cal O},{\cal O}']}(n)\, \Delta_\delta^{2}(q)\, P(q)\ .
\eeq
In the real universe, the two functions ${\cal I}^{[\delta^2,\G_2]}$ and ${\cal I}^{[\G_2,\G_2]}$ are approximately proportional to each other,
 \beq
 {\cal I}^{[\delta^2,\G_2]}(q)\approx -\frac{7}{5}{\cal I}^{[\G_2,\G_2]}(q) \ ,
 \eeq
 as illustrated in fig.~\ref{fig:Phh}.
The function ${\cal I}^{[\delta^2,\delta^2]}$ is divergent, but the finite piece can be evaluated by subtracting the $q=0$ divergent part, 
\beq
\widehat {\cal I}^{[\delta^2,\delta^2]}(q) \equiv {\cal I}^{[\delta^2,\delta^2]}(q)-{\cal I}^{[\delta^2,\delta^2]}(0)=2\int_{\p}P(p)\big(P(|\q-\p|)-P(p)\big)\ .
\eeq
This expression does not remove (possibly divergent) contact terms of the form $q^{2n}$ for integer $n\geq 1$, which are derivatives of delta functions in position space.  These terms can be removed by hand without altering correlation functions at finite separation.  This freedom to alter contact terms can be made explicit by introducing a stochastic bias parameter (see e.g.~\cite{McDonald:2006mx}).  We will simply ignore them with the understanding that our expressions are correct up to such contact terms.  For the real universe, these contact terms are known to be highly suppressed in the quasi-linear regime~\cite{Baumann:2010tm, Carrasco:2012cv}.

\subsection{Summary: Effective Bias Parameters}
\label{sec:summary1}

\begin{figure}[ht]
   \centering
       \includegraphics[width=0.8\textwidth]{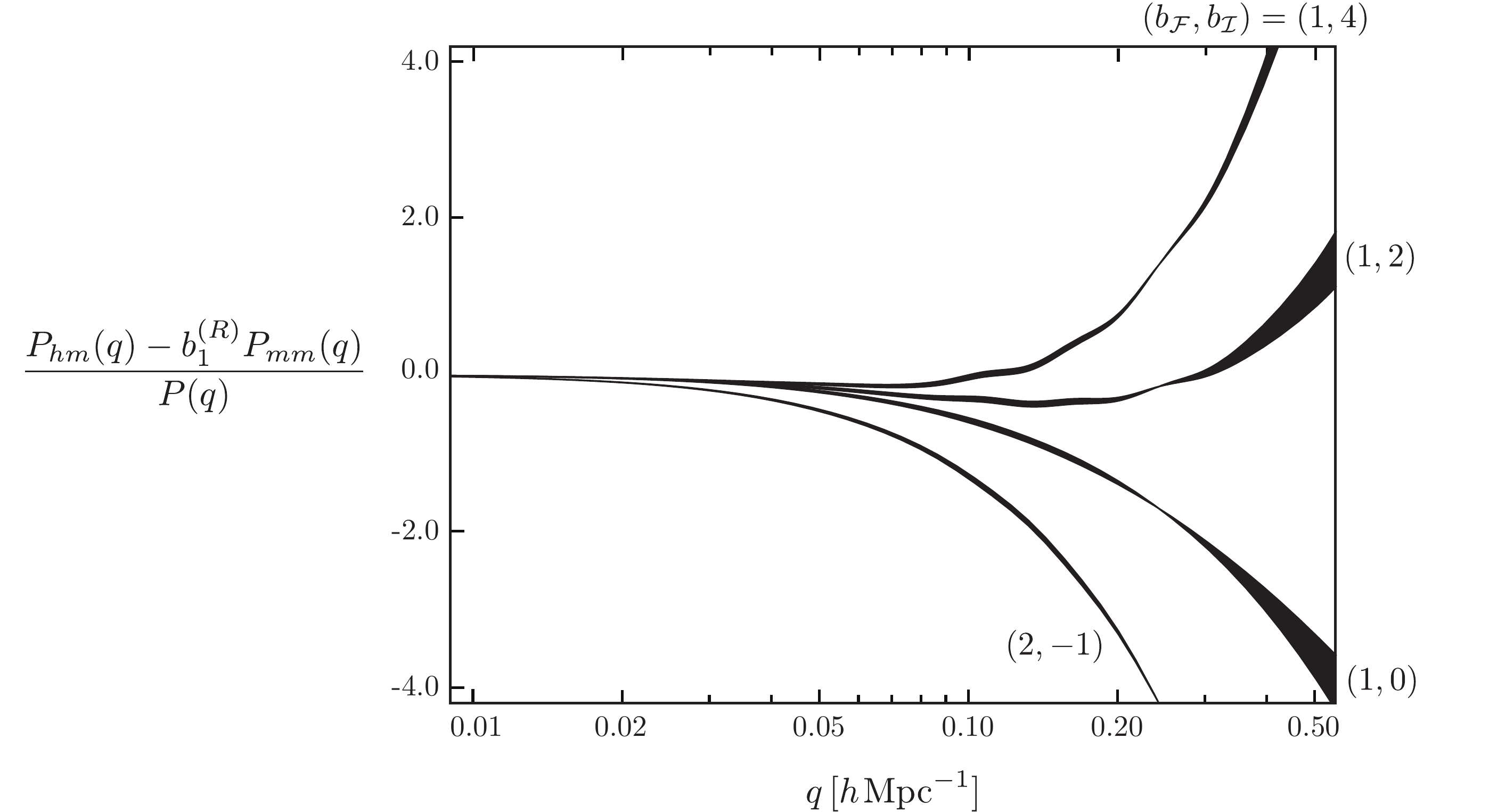}
    \caption{Illustration of the degeneracy of the different contributions to the halo-matter power spectrum. For each curve, $b_{\cal F}$ and $b_{\cal I}$ are kept fixed, but $b_{\G_2}^{(R)}$ is varied between $-3$ and $+3$. }
  \label{fig:Phm-Degeneracy}
\end{figure}

\begin{figure}[h!]
   \centering
       \includegraphics[width=0.8\textwidth]{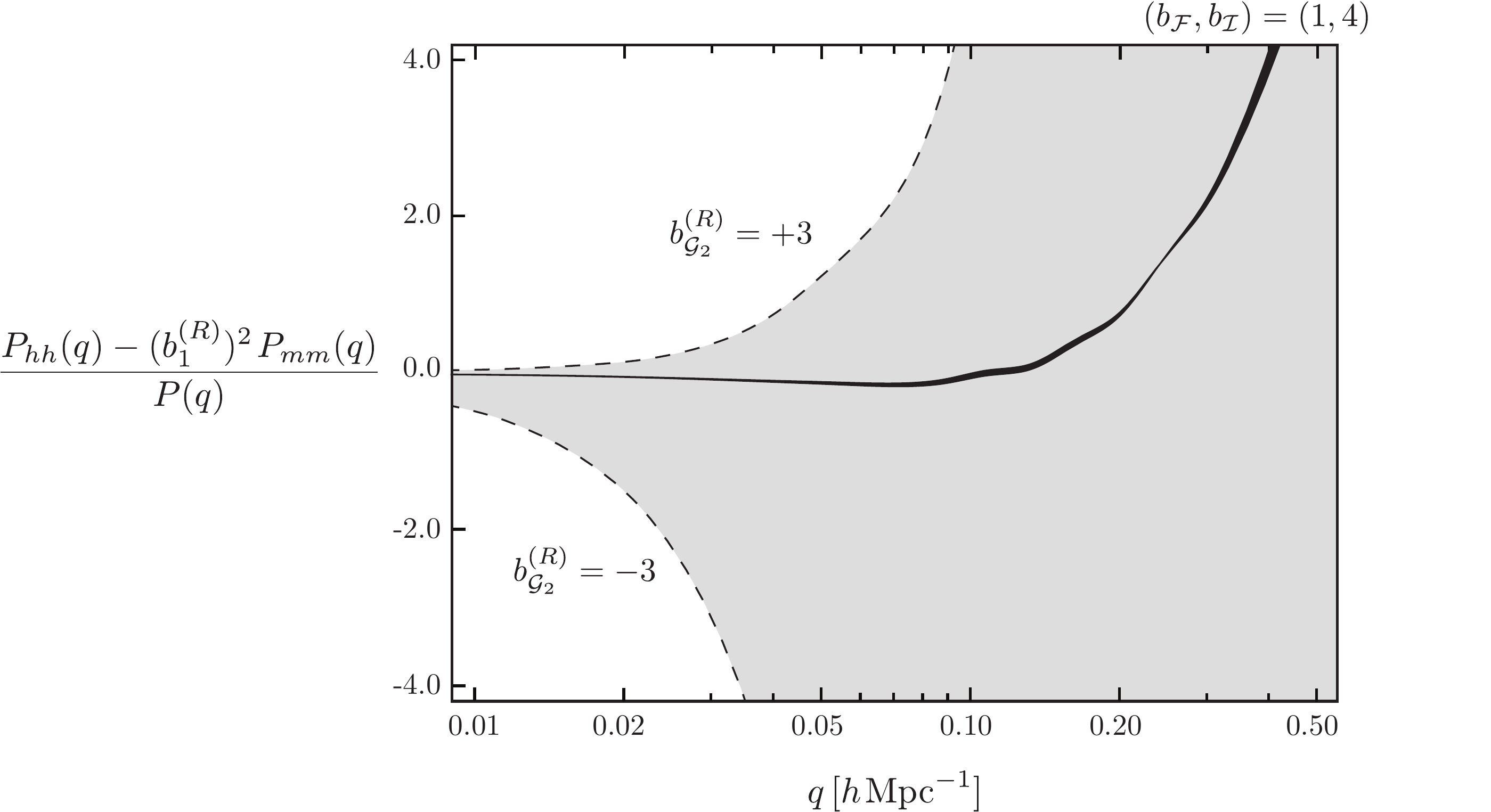}
    \caption{Illustration of how the degeneracy presented in fig.~\ref{fig:Phm-Degeneracy} is broken by the halo-halo power spectrum. While $P_{hm}$ (black band) mostly depends only on the effective bias parameters $b_{\cal F}$ and $b_{\cal I}$, $P_{hh}$ (gray band) is sensitive to $b_{\G_2}^{(R)}$.}
  \label{fig:Phh-Degeneracy}
\end{figure}

We have arrived at the following predictions for the halo power spectra 
\begin{align}
P_{hm}(q) &= b_L^{(R)}(q) P_{mm}(q)    +  \sum_{{\cal O} } b_{{\cal O}}^{(R)}\,  {\cal I}^{[{\cal O}]}(q)  \ , \label{equ:Phm2} \\
P_{hh}(q) &= b_1^{(R)} \left[b_1^{(R)}P_{mm}(q)+2\sum_{\cal O}b_{\cal O}^{(R)}\left({\cal F}^{[{\cal O}]}(q)+{\cal I}^{[{\cal O}]}(q)\right)\right] + \sum_{{\cal O}, {\cal O}'}b_{{\cal O}}^{(R)} b_{{\cal O}'}^{(R)}\, {\cal I}^{[{\cal O},{\cal O}']}(q)\ , \label{equ:Phh2}
\end{align}
where $b_L^{(R)}$ includes contributions from $\delta$, $\G_2$ and $\Gamma_3$, while the sums over ${\cal I}$-terms are only over the operators $\delta^2$ and $\G_2$.
Fitting (\ref{equ:Phm2}) and (\ref{equ:Phh2}) to data over a sufficiently wide range of momenta, in principle, allows the effective bias parameters $b_1^{(R)}$, $b_{2}^{(R)}$,
$b_{\G_2}^{(R)}$ and $b_{\Gamma_3}^{(R)}$ to be determined.  In practice, extracting all the biasing coefficients from the power spectra alone can be challenging (if not impossible).  For example,  in a scaling universe, the functions ${\cal F}(q)$, ${\cal I}^{[{\cal O}]}(q) $ and ${\cal I}^{[{\cal O},{\cal O}']}(q)$ are identical powers of $q$ for $b_{2}^{(R)}$, $b_{\G_2}^{(R)}$ and $b_{\Gamma_3}^{(R)}$.  
In that case, measurements of the power spectra only determine two linear combinations of these three bias coefficients. The real universe is sufficiently close to a scaling universe (in the regimes of interest), that these problems may persist. 
In fact, taking into account the near-degeneracy of the ${\cal I}$-terms, we can write the halo-matter power spectrum as
\beq
\frac{P_{hm}(q)-b_1^{(R)}P_{mm}(q)}{P(q)} \approx   b_{\cal F}\hskip 1pt f(q) + b_{\cal I} \hskip 1pt i(q)\ ,
\eeq
where $i(q) \equiv {\cal I}(q)/P(q)$ and
\beq
b_{\cal F}\equiv b_{{\cal G}_2}^{(R)}+\frac{2}{5}b_{\Gamma_3}^{(R)}\qquad{\rm and}\qquad b_{\cal I}\equiv b_{\delta^2}^{(R)}- \frac{4}{5}\hskip 2pt b_{\G_2}^{(R)}\ .
\eeq 
The functions $f(q)$ and $i(q)$ are sufficiently different that it should be possible to measure the effective bias parameters $b_{\cal F}$ and $b_{\cal I}$ independently. However, a degeneracy between the parameters $b_{2}^{(R)} \equiv \frac{1}{2}b_{\delta^2}^{(R)}$,
$b_{\G_2}^{(R)}$ and $b_{\Gamma_3}^{(R)}$ does remain, see fig.~\ref{fig:Phm-Degeneracy}. 
This degeneracy may be broken by  considering the halo-halo power spectrum,
\begin{align}
\frac{P_{hh}(q)}{P(q)} &\approx b_1^{(R)} \left[b_1^{(R)}\frac{P_{mm}(q)}{P(q)} +2\big(b_{\cal F}\hskip 1pt f(q)+b_{\cal I}\hskip 1pt i(q)\big)\right]\nonumber\\
&\hskip 40pt + \left(b_{\cal I}+\frac{4}{5}\hskip 2pt b^{(R)}_{{\cal G}_2}\right)^2\,  {i}^{[\delta^2,\delta^2]}(q)+ \frac{5}{7}\left[\big(b_{{\cal G}_2}^{(R)}\big)^2 - \frac{1}{2} b_{\cal I}\hskip 1pt b_{{\cal G}_2}^{(R)}\right] i^{[\G_2,\G_2]}(q)\ .
\end{align}
We see that if $b_1^{(R)}$, $b_{\cal F}$ and $b_{\cal I}$ are determined from the halo-matter power spectrum, the halo-halo power spectrum allows a measurement of $b_{\G_2}^{(R)}$ (and hence also determines $b_{\delta^2}^{(R)}$ and $b_{\Gamma_3}^{(R)}$). These considerations are illustrated in fig.~\ref{fig:Phh-Degeneracy}.


\subsection{Higher-Point Statistics}
\label{sec:HPF}

An alternative to break the degeneracy between the different bias contributions is to consult higher-point statistics.
In this section, we present an analysis of the bispectrum
\beq
B_{hmm}(q,q_1,q_2) \equiv\vev{(\delta_h)_{\q}\hskip 1pt\delta_{\q_1}\delta_{\q_2}}'\ , \label{equ:Bhmm}
\eeq
and the trispectrum
\beq
T_{hmmm}(q,q_1,q_2,q_3) \equiv\vev{(\delta_h)_\q\hskip 1pt\delta_{\q_1}\delta_{\q_2}\delta_{\q_3}}'\ . \label{equ:Thmmm}
\eeq 
The bispectrum will be evaluated at one loop, while the trispectrum will only be presented at tree level. Details of the bispectrum calculation can be found in Appendix~\ref{app:Bispectrum}.

\subsubsection{Three-Point Function}

At tree level, the bispectrum (\ref{equ:Bhmm}) gets contributions from $\delta$, $\delta^2$ and $\G_2$: 
\begin{align}
B_{hmm}(q,q_1,q_2)	&\ =\ b_1^{(R)}B_{mmm}(q,q_1,q_2) +\Big(b_{2}^{(R)}+2b_{\G_2}^{(R)}\big(\mu_{12}^2-1\big)\Big)P_{mm}(q_1)P_{mm}(q_2) \ ,\label{equ:Bhmm2}
\end{align}
where $\mu_{12}\equiv \q_1 \hskip -1pt\cdot \hskip -1pt\q_2/q_1q_2$ and $B_{mmm}(q,q_1,q_2) \equiv \vev{\delta_{\q} \hskip 1pt \delta_{\q_1}\delta_{\q_2}}'$ is the dark matter bispectrum. 
We see that the contributions from $\delta^2$ and $\G_2$ come 
 with distinct momentum dependences, which has been used previously to fit for the parameters $b_{2}^{(R)}$ and $b_{\G_2}^{(R)}$ in N-body simulations~\cite{Chan:2012jj,Baldauf:2012hs}.  However, we note that knowledge of $b_{2}^{(R)}$ and $b_{\G_2}^{(R)}$ is not sufficient to check the consistency with the one-loop power spectrum, which includes the additional bias parameter $b_{\Gamma_3}^{(R)}$. 
 
 \vskip 4pt
At one-loop, operators up to quartic order contribute to (\ref{equ:Bhmm}). 
However, composite operators with four legs, such as $\delta^4$, do not contribute after renormalization. We therefore only have to consider operators with up to three legs.  In total there are eleven such operators, including all the operators in (\ref{equ:OP}). These operators are discussed in detail in Appendix~\ref{app:Bispectrum}. Here, we collectively call them ${\cal O}$.  The one-loop diagrams can be organized into two classes of contributions: 
\begin{itemize}
\item {${\cal F}${\it-terms.}}---Diagrams whose only contractions are between the internal legs of the composite operator are proportional to the dark matter power spectra 
\beq
\langle [{\cal O}]_{\q}\hskip 1pt \delta_{\q_1} \delta_{\q_2} \rangle' \, \subset\,  {\cal F}_B^{[{\cal O}]}(\q_1,\q_2) = f_B^{[{\cal O}]}(\q_1,\q_2) P_1 P_2 \ ,
\eeq
where the functions $f_B^{[{\cal O}]}(\q_1,\q_2)$ are defined explicitly in Appendix~\ref{app:Bispectrum}. 
\item {${\cal I}${\it-terms.}}---Diagrams with at least one contraction with an external leg contain non-trivial convolutions between the dark matter power spectra and the kernel functions of standard perturbation theory, 
\beq
\hspace{-3.33cm}\langle [{\cal O}]_{\q}\hskip 1pt \delta_{\q_1} \delta_{\q_2} \rangle' \, \subset\,  {\cal I}_B^{[{\cal O}]}(\q_1,\q_2)  \ .
\eeq
Six of the eleven operators have such contributions, which are presented explicitly in Appendix~\ref{app:Bispectrum}.
\end{itemize}
In the end, the one-loop bispectrum can be written as
 \beq
\frac{B_{hmm} - b_1^{(R)}B_{mmm}}{P_1 P_2}
= \left( b_2^{(R)}+ 2b_{\G_2}^{(R)}\big(\mu_{12}^2-1\big) + \sum_{\cal O} b_{{\cal O}}^{(R)} f_B^{[{\cal O}]}\right) +\, \sum_{{\cal O}} b_{{\cal O}}^{(R)}\frac{ {\cal I}^{[{\cal O}]}_B}{P_1 P_2}  \ . \label{equ:B1loop}
\eeq
Due to the complexity of the final answer, the one-loop contributions to the bispectrum are probably of limited use in determining the effective bias parameters. 
However, knowledge of the functional form of the one-loop corrections gives us a handle on the expected theoretical error in the tree-level ansatz~(\ref{equ:Bhmm2}).

\subsubsection{Four-Point Function}
\label{sec:4pt}

We have seen that the one-loop power spectra in \S\ref{sec:2point} involve contributions proportional to~$b_{\Gamma_3}^{(R)}$.
This parameter can be measured from the tree-level trispectrum~(\ref{equ:Thmmm}), which receives contributions from every operator up to cubic order
\beq
T_{hmmm}(q,q_1,q_2,q_3)\, =\, b_1^{(R)}\,T_{mmmm}(q,q_1,q_2, q_3)  + \sum_{{\cal O}\in\{{\cal O}_{2},{\cal O}_{3}\}}\hskip -4pt b_{[{\cal O}]}^{(R)}\vev{[{\cal O}]_{\q}\hskip 1pt\delta_{\q_1}\delta_{\q_2}\delta_{\q_3}}'\ ,\label{equ:trisp}
\eeq
where ${\cal O}_{2}\equiv\{[\delta^2]\hskip1pt,\hskip1pt[\G_2]\}$, ${\cal O}_{3}\equiv\{[\delta^3]\hskip1pt,\hskip1pt[\G_2\delta]\hskip1pt,\hskip1pt[\G_3]\hskip1pt,\hskip1pt[\Gamma_3]\}$, and $T_{mmmm}(q,q_1,q_2, q_3)  \equiv \vev{\delta_{\q} \hskip 1pt\delta_{\q_1}\delta_{\q_2}\delta_{\q_3}}'$. 
If the linear and quadratic bias parameters are measured from the power spectrum and the bispectrum, then the trispectrum can be used to extract the cubic bias parameters~${\cal O}_{3}$.
At leading order in perturbation theory, we have
\beq
\vev{[{\cal O}_{3}]_{\q}\hskip 1pt \delta_{\q_1}\delta_{\q_2}\delta_{\q_3}}^\prime = 6\hskip 1pt {g}^{[{\cal O}_{3}]}P_1P_2P_3\ ,
\eeq
where the functions $g^{{[{\cal O}_{3}]}}$ contain the specific momentum dependence associated with each operator in ${\cal O}_{3}$: 
\begin{align}
g^{[\delta^3]} & = 1 \ , \\
g^{[\G_2 \delta]} &= \frac{1}{3} \left( \sigma_{12}^2+\sigma_{23}^2+\sigma_{13}^2 \right)\ , \\
g^{[\G_3]} &=\frac{1}{2}\left[\left(\mu_{12}^2+\mu_{23}^2+\mu_{13}^2 \right)- 2\hskip 1pt\mu_{12}\mu_{23}\mu_{13}-1\right] \ ,\\
 g^{[\Gamma_3]} &= - \frac{4}{21}\left( \sigma_{1,23}^2  \sigma_{23}^2 +2\,{\rm perms}.\right) \ .
\end{align}
We see that the different contributions can be distinguish on the basis of their unique momentum dependences in the trispectrum.
In particular, it is possible to extract information about the velocity tidal tensor $\Gamma_3(\Phi_g,\Phi_v)$ and measure the associated bias parameter~$b_{\Gamma_3}^{(R)}$.

\subsection{Summary: Halo Statistics}
\label{sec:summary}

The following collects our results for the halo-matter power spectrum, eq.~(\ref{equ:Phm2}), the bispectrum, eq.~(\ref{equ:B1loop}), and trispectrum, eq.~(\ref{equ:trisp}):
\begin{align}
\frac{P_{hm}(q) - b_1^{(R)} P_{mm}(q)}{P(q)} &\, =\,   \Big(b_{\G_2}^{(R)} + \frac{2}{5}b_{\Gamma_3}^{(R)} \Big) f(q)    +  \sum_{{\cal O} } b_{{\cal O}}^{(R)}\,  \frac{{\cal I}^{[{\cal O}]}(q)}{P(q)}  \ , \\
\frac{B_{hmm} - b_1^{(R)}B_{mmm}}{P_1 P_2}
&\, =\,\Big( b_{2}^{(R)}+ 2b_{\G_2}^{(R)}\big(\mu_{12}^2-1\big)\Big) + \sum_{\cal O} b_{{\cal O}}^{(R)} f_B^{[{\cal O}]} \, +\, \sum_{{\cal O}} b_{{\cal O}}^{(R)}\frac{ {\cal I}^{[{\cal O}]}_B}{P_1 P_2} \ , \\
\frac{T_{hmmm} - b_1^{(R)}T_{mmmm}}{6 P_1 P_2 P_3}
&\, =\,   \sum_{{\cal O}} b_{{\cal O}}^{(R)} g^{[{\cal O}]} + ({\it loops})\ ,
\end{align}
where the functions $f$, $g$ and ${\cal I}$ are defined in \S\ref{sec:2point}, \S\ref{sec:HPF} and Appendix~\ref{app:Bispectrum}. Table~\ref{Table:Summary} summarizes how the effective bias parameters appear in these results. This shows that the parameters $b_1^{(R)}$, $b_{2}^{(R)}$, $b_{\G_2}^{(R)}$ and $b_{\Gamma_3}^{(R)}$ can be extracted by fitting to the tree-level results. The one-loop contributions to the power spectrum then become predictions.

\vspace{0.5cm}

	 \begin{table}[h!]

	\heavyrulewidth=.08em
	\lightrulewidth=.05em
	\cmidrulewidth=.03em
	\belowrulesep=.65ex
	\belowbottomsep=0pt
	\aboverulesep=.4ex
	\abovetopsep=0pt
	\cmidrulesep=\doublerulesep
	\cmidrulekern=.5em
	\defaultaddspace=.5em
	\renewcommand{\arraystretch}{1.6}

	\begin{center}
		\small
		\begin{tabular}{llll}

			\toprule
		&\multicolumn{1}{c}{Tree-Level} &\multicolumn{2}{c}{One-Loop} \\[-4pt]
		&\multicolumn{1}{c}{} &\multicolumn{1}{c}{${\cal F}$-terms} &\multicolumn{1}{c}{${\cal I}$-terms} \\[2pt]
			\midrule
		\rowcolor[gray]{0.9}{} Power Spectrum\ \ \ \ & $b_1^{(R)}$  &  	&			\\[2pt]
		\rowcolor[gray]{0.9}{} &  &$b_{{\cal G}_2}^{(R)}$ & $b_{2}^{(R)}$, $b_{{\cal G}_2}^{(R)}$\\[2pt]
		\rowcolor[gray]{0.9}{} &  &$b_{{\Gamma}_3}^{(R)}$ &			\\[2pt]
		\midrule
		\rowcolor[gray]{0.9}{} Bispectrum &$b_1^{(R)}$ &  &				\\[2pt]
		\rowcolor[gray]{0.9}{}  &$b_{2}^{(R)}$, $b_{{\cal G}_2}^{(R)}$ & $b_{2}^{(R)}$\hskip 2pt, $b_{{\cal G}_2}^{(R)}$  & $b_{2}^{(R)}$\hskip 2pt, $b_{{\cal G}_2}^{(R)}$				\\[2pt]
		\rowcolor[gray]{0.9}{}  & & $b_{3}^{(R)}$\hskip 2pt, $b_{{\cal G}_2\delta}^{(R)}$\hskip 2pt, $b_{{\cal G}_3}^{(R)}$\hskip 2pt, $b_{{\Gamma}_3}^{(R)}$ \ \   &	$b_{3}^{(R)}$\hskip 2pt, $b_{{\cal G}_2\delta}^{(R)}$\hskip 2pt, $b_{{\cal G}_3}^{(R)}$\hskip 2pt, $b_{{\Gamma}_3}^{(R)}$	\ \ 		\\[2pt]
		\rowcolor[gray]{0.9}{}  & & $b_{\Gamma_4}^{(R)}$\hskip 2pt, $b_{\tilde\Gamma_4}^{(R)}$\hskip 2pt, $b_{\Gamma_3 \delta}^{(R)}$\hskip 2pt, $b_{\Delta_4}^{(R)}$ &		\\[3pt]
		\midrule
		\rowcolor[gray]{0.9}{} Trispectrum&$b_1^{(R)}$  &  &			\\[2pt]
		\rowcolor[gray]{0.9}{}  &$b_{2}^{(R)}$\hskip 2pt, $b_{{\cal G}_2}^{(R)}$ &\multicolumn{1}{l}{\cellcolor[gray]{0.9}{}  {\it many} } &\multicolumn{1}{l}{\cellcolor[gray]{0.9}{}  {\it many} }	\\[2pt]
		\rowcolor[gray]{0.9}{}  &  $b_3^{(R)}$\hskip 2pt, $b_{{\cal G}_2\delta}^{(R)}$\hskip 2pt, $b_{{\cal G}_3}^{(R)}$\hskip 2pt, $b_{{\Gamma}_3}^{(R)}$ \ \  &  &		\\[2pt]
		\bottomrule
		\end{tabular}
	\end{center}
	\vspace{-0.2cm}
	\caption{Summary of the dependence of the halo statistics on the renormalized bias parameters.}
	\label{Table:Summary}
	\end{table}

\section{Conclusions}
\label{sec:Conclusions}

In this paper, we have shown explicitly how renormalization 
{\it forces} us to treat biasing models as {\it effective theories}, i.e.~as a double expansion in terms of fluctuations in the dark matter density and velocity and derivatives thereof.
Consistently removing the short-scale physics in the local Eulerian biasing model~\cite{Fry:1992vr} doesn't just renormalize the bias parameters, but also generates {non-local} terms and {higher-derivative} contributions.  In order for the theory to become independent of the unphysical regulator of composite operators, all terms consistent with the symmetries have to be included in the biasing model. At lowest order, this means adding the gravitational tidal tensor, while at cubic order, the {velocity potential} appears as an independent degree of freedom.

In the process, we have clarified a few technical aspects of the renormalization procedure proposed by McDonald~\cite{McDonald:2006mx}.  
We organized the renormalization of composite operators in a convenient diagrammatic representation and derived the building blocks for the one-loop renormalization of the halo power spectrum and bispectrum.  
We proved that Galileon operators aren't renormalized at leading order in derivatives. 
Finally, we showed explicitly how the definition of the renormalized theory depends on the renormalization scale and how the terms in the bias expansion mix as this scale is varied.  This scale-dependence of the renormalized halo bias is relevant for interpreting recent N-body results~\cite{Chan:2012jx, Sheth:2012fc, Pollack:2013alj}.

\vskip 4pt
Our work motivates several future directions:

\begin{itemize}
\item Most importantly, it remains to be quantified how many terms need to be kept in the effective theory to achieve a given target of precision in the predictions for the statistics of halos.
The importance of including the gravitational tidal tensor has been established through N-body simulations in \cite{Baldauf:2012hs,Chan:2012jj}.  However, a completely systematic exploration of the effective theory of halo biasing has not yet been performed.
For example, the scale $\Lambda_\star$ which suppresses higher-derivative terms has not been measured and the errors that arise from truncating the effective theory have not been quantified.

\item A key motivation for understanding non-linear biasing is primordial non-Gaussianity.
One may hope that the non-linearities arising from the biasing are sufficiently distinct, so that the primordial signals can be extracted from the shape information of the correlation functions.  To analyze this self-consistently, we must consider the effects of these non-Gaussian contributions on the renormalization of the biasing model~\cite{McDonald:2008sc, NG-progress}.  

\item We have focused on Eulerian biasing, but it has been suggested that biasing in Lagrangian space may have some advantages.  The approach taken here can likely be adapted to the Lagrangian EFT-of-LSS \cite{Porto:2013qua}.  Which scheme is more useful may depend on the type of observable that is considered.

\item  We have shown how renormalization forces biasing to be non-local in space, but the theory has remained local in time. Recently, the necessity of non-locality in time has been emphasized in the EFT-of-LSS~\cite{Carrasco:2013sva, Carroll:2013oxa}.  
This deserves further consideration in the present context.

\item Ultimately, galaxies are observed in redshift space. Redshift space distortions can add extra non-linear contributions to the observed galaxy correlation functions. This should be taken into account.

\item  We have seen that a biasing model in terms of just the dark matter density  is not consistent and that the velocity potential has to be added as independent degree of freedom.   It would be interesting to explore the physical effects of the velocity potential more widely, e.g.~beyond perturbation theory. 
\end{itemize}

\subsubsection*{Acknowledgements}

We thank Tobias Baldauf, Hayden Lee, Enrico Pajer, Marcel Schmittfull, Leonardo Senatore and Marko Simonovi\'c for helpful discussions. 
D.B.~and V.A.~gratefully acknowledge support from a Starting Grant of the European Research Council (ERC STG grant 279617).  The research of D.G.~is supported in part by the Stanford Institute for Theoretical Physics and by the U.S. Department of Energy contract to SLAC no.\ DE-AC02-76SF00515.
M.Z.~is supported in part by the NSF grants PHY-0855425, AST-0907969, PHY-1213563 and by the David and Lucile Packard Foundation.

\newpage
\appendix

\section{Scale Dependence and Operator Mixing}
\label{sec:mixing}

In the Wilsonian renormalization scheme that we have adopted in this paper, the bare bias parameter will depend on the cutoff,\footnote{It would be interesting to relate this $\Lambda$-dependence to the dependence of scatter-plot bias parameters on the smoothing scale that is observed in N-body simulations.} $b_{\cal O}^{(0)}(\Lambda)$, in a way that is dictated by the renormalization group equations
\beq
\frac{d b_{\cal O}^{(R)}}{d \Lambda} = 0\ .
\eeq
Although the $\Lambda$-dependence of bare quantities isn't physical, it is often an indication that the corresponding physical quantities depend on the scale at which they are measured.  This is typically reflected in the Callan-Symanzik equation applied to  correlation functions.  In this appendix, we will show that this intuition is correct, although the use of the Callan-Symanzik equation will not be necessary.\footnote{Renormalization group flow is usually most useful for logarithmic divergences.  For a scaling universe with $P(q) \propto q^n$, the variance $\sigma^2(\Lambda)$ is logarithmically divergent only for $n=-3$, in which case there are also infrared divergences.  For this reason, we will find this language less useful.}

\vskip 4pt
In Section~\ref{sec:Renorm2}, we defined the renormalization conditions in the long-wavelength limit ($q \to 0$), where non-linearities are negligible~($\delta \to \delta^{(1)}$).  At finite separation,\footnote{We chose to work in position space to avoid contributions that are delta-function localized.  However, this means that our definition of $\delta_h$ is only correct up to contact terms.  The above procedure can also be applied in momentum space with $q = \mu$, if one is careful to identify only the terms that are non-analytic in $q$.}  $|\x| = 2\pi/\mu$,  the basis of renormalized operators, $[{\cal O}_I]$, will therefore, in general, not be orthogonal
\beq
\langle [{\cal O}_I](\x) [{\cal O}_J]({\bf 0}) \rangle \big|_{|\x| = 2\pi/\mu} \,=\, \Gamma_{IJ}(\mu, \knl) \ , 
\eeq
where $\Gamma_{IJ}(\mu, \knl) $ is a real positive-definite matrix.\footnote{Here, we are ignoring ``descendants" of ${\cal O}$, i.e.~operators which are total derivatives of ${\cal O}$.  Since these terms are fully correlated with ${\cal O}$, they cannot be diagonalized.}  
Of course, we can diagonalize $\Gamma_{IJ}$ in terms of a new basis of operators, 
\beq
[\widehat {\cal O}_I](\mu) = M_I{}^J(\mu) [{\cal O}_J]\ ,
\eeq
 such that $\langle [\widehat{\cal O}_I](\x) [\widehat{\cal O}_J]({\bf 0})\rangle \big|_{x = 2\pi/\mu} \,\propto\, \delta_{IJ}$.
The basis of operators that is orthogonal at a scale $\mu$ will mix and in general won't be an orthogonal basis at another scale $\mu'$.
 In order for the halo density contrast to be independent of $\mu$, the renormalized bias parameters need to be $\mu$-dependent and the bias expansion is
\beq\label{equ:muhalo}
\delta_h \, =\, \sum_I\, \widehat b_I^{(R)}(\mu)\, [\widehat {\cal O}_I](\mu)\ .
\eeq
The orthogonal basis $[\widehat {\cal O}_I]$ is convenient because it, in principle, allows us to determine\footnote{A similar approach to measuring bias parameters was taken in \cite{Pollack:2013alj}, although it wasn't applied to the renormalized theory.} all bias coefficients unambiguously:
\beq\label{equ:mubias}
\widehat b_I^{(R)}(\mu) \equiv \frac{ \langle \delta_h{}(\x)  [\widehat {\cal O}_I]({\bf 0}) \rangle' }{ \langle [\widehat{\cal O}_I](\x) [\widehat {\cal O}_I]({\bf 0})\rangle'} \Bigg|_{|\x|=2\pi/\mu}\ .
\eeq

  What are we to make of the $\mu$-dependence of the bias parameters?
Let us first consider the case where the correlation functions of $\delta$ are purely Gaussian.  In this case, we can define $[\delta^n] = H_n (\delta)$~\cite{Baumann:2012bc}, where $H_n$ are Hermite polynomials, such that 
\beq
\langle [\delta^n](\x) [\delta^m] ({\bf 0}) \rangle' =n!\, \xi^{n}(|\x|) \, \delta_{nm} \ ,
\eeq
where $\xi(|\x|)\equiv \vev{\delta^{(1)}(\x) \delta^{(1)}({\bf 0})}'$.
In terms of the Hermite polynomials, the bias coefficients $\widehat b_n^{(R)}$ are independent of $\mu$.  This basis is diagonal at all scales.  The $\mu$-dependence of the bias coefficients must therefore arise from non-linear evolution.

As a simple example, let us consider the mixing between the operators $[\delta]$ and $[\delta^2]$ as a function of separation. 
 The detailed form of this mixing was computed in \S\ref{sec:delta2}.
  In a scaling universe, where $P \propto q^n$, dimensional analysis shows that to fourth order in $\delta^{(1)}$, we have
\beq
\Gamma = P(\mu) \left( \begin{array}{cc}
 1 + \alpha(n)\varepsilon &\ \beta(n) \varepsilon\  \\
\beta(n)\varepsilon & \ \gamma(n) \varepsilon\ \end{array} \right)\ ,
\eeq
where $\varepsilon \equiv (\mu/\knl)^{3+n}$.  To order $\varepsilon^2$, the orthogonal basis of operators is 
\begin{align}
\widehat{\cal O}_1 &\equiv [\delta] + \beta \varepsilon [\delta^2]\ , \label{equ:hatO1}\\ 
\widehat{\cal O}_2 &\equiv [\delta^2]- \beta \varepsilon  [\delta]\ . \label{equ:hatO2}
\end{align} 
As a result, 
we get the following relationship between the bias parameters measured at $\mu$ and $\mu'$ (dropping mixing with higher-derivative terms)
 \begin{align}
 \widehat b_1^{(R)}(\mu') &\, =\, \widehat b_1^{(R)}(\mu) - \beta(n) (\varepsilon'-\varepsilon)\,   \widehat b_2^{(R)}(\mu) \ , \\
 \widehat b_2^{(R)}(\mu') &\,=\, \widehat b_2^{(R)}(\mu) + \beta(n) (\varepsilon'-\varepsilon)\, \widehat   b_1^{(R)}(\mu) \ .
\end{align}
We see that for $n \gg -3$, the scale dependence of the bias parameters vanishes rapidly as we approach the linear regime, $\mu \ll \knl$. 
Of course, there can still be important mixing with higher-derivative terms even for large values of $n$. 
  However, the measurement of higher-derivative bias terms is more complicated as the bias parameters cannot always be diagonalized.

\newpage
\section{Renormalization of Higher-Order Operators}
\label{ap:HOO}

In this appendix, we describe in more detail the renormalization of operators of third order and higher.

\subsection{Renormalization of $\delta^n $}
\label{ssec:HOO}

First, we show that the one-loop renormalization of $\delta^n$, with $n>2$, is completely determined by the one-loop renormalization of $\delta^2$. More precisely,  at one-loop, we have
\beq
[\delta^n]^{\rm loop}\ =\ \frac{n(n-1)}{2}\,[\delta^2]^{\rm loop}\,[\delta^{n-2}]^{\rm tree}\ ,\label{eq:deltan1loop}
\eeq
where the superscript ``loop'' indicates that two (and only two) linear $\delta^{(1)}$'s are contracted inside of $\delta^n$, while the superscript ``tree'' denotes that the $\delta$'s of $\delta^{n-2}$ are contracted only with the external legs $\delta^{(1)}$.

The proof of the statement (\ref{eq:deltan1loop}) is quite straightforward.  Consider the correlation of $\delta^n$ with $E$ external legs. There are $\frac{n(n-1)}{2}$ ways of contracting two of the $\delta$'s in $\delta^n$ to form a loop. The external legs are then contracted with either the $\delta^2$ forming the loop or with the $n-2$ remaining $\delta$'s. The following is a diagram in which $r\leq E$ external legs are contracted with $\delta^2$, while the rest are contracted with $\delta^{n-2}$\hskip 2pt:
\begin{figure}[h!]
   \centering
       \includegraphics[scale =1.0]{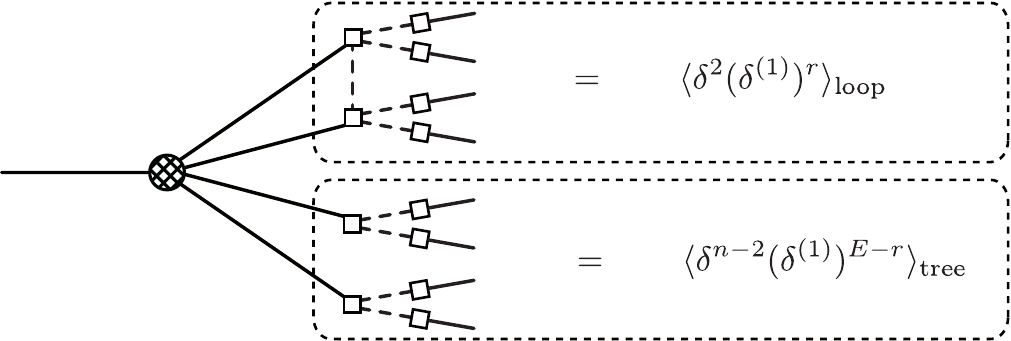}
  \label{fig:deltan}
\end{figure}

\noindent
The full correlation function is then obtained by summing over $r$\hskip 2pt:
\beq
\vev{\delta^n\hskip 1pt (\delta^{(1)})^E}_{p\hskip 1pt{\rm 1PI}}\ =\ \frac{n(n-1)}{2}\sum_{r=0}^E\Big(\vev{\delta^2 \hskip 1pt (\delta^{(1)})^r}_{{\rm 1PI}} \hskip 1pt \vev{\delta^{n-2}\hskip 1pt(\delta^{(1)})^{E-r}}_{\rm tree}\ + {\rm perms}\Big)\ ,
\eeq
which implies~(\ref{eq:deltan1loop}). 

This result relates (at one loop) the counterterms of $\delta^2$ and those of $\delta^n$
\beq
\sum_{\widetilde{\cal O }}Z^{(\delta^n)}_{\widetilde{\cal O}}\widetilde{\cal O}\ =\ \frac{n(n-1)}{2}\, \delta^{n-2}\, \sum_{\widetilde{\cal O} }Z^{(\delta^2)}_{ \widetilde{\cal O} }\hskip 1pt \widetilde{\cal O} \ .
\eeq
For example, we can infer the one-loop counterterms for the renormalization of~$\delta^3$ and $\delta^4$ directly from (\ref{eq:ct2}),
\begin{align}
 \sum_{\widetilde{\cal O}}Z_{\widetilde{\cal O}}^{(\delta^3)}{\widetilde{\cal O}} &\ =\ -3\hskip 1pt \sigma^2(\Lambda)\left[\, \delta + \frac{68}{21}\, \delta^2 \, \right] \ , \\
 \sum_{\widetilde{\cal O}}Z_{\widetilde{\cal O}}^{(\delta^4)}{\widetilde{\cal O}}  &\ =\ - 6 \hskip 1pt \sigma^2(\Lambda) \hskip 1pt \delta^2\ ,
\end{align}
where we have only shown terms up to order $(\delta^{(1)})^4$.

\subsection{Renormalization of $\G_2\delta$}

In the main text, we argued that the one-loop renormalization of the linear bias requires the renormalization of all cubic operators. However, the renormalization of the operator $\delta^3$ is related to the renormalization of $\delta^2$ (see \S\ref{ssec:HOO}) and the Galileon operators aren't renormalized at leading order in derivatives (see Appendix~\ref{app:Galileon}),  so the only operator which remains to be renormalized explicitly is  $\G_2\hskip 1pt \delta$. Since the renormalization of this operator is identical to the renormalization of $\delta^2$, we just state the final result 
\begin{align}
\sum_{\widetilde{\cal O}}Z^{(\G_2 \delta)}_{\widetilde{\cal O}} \widetilde{\cal O}&\ =\ \frac{4}{3}\sigma^2(\Lambda)\left[ \, \delta+\frac{94}{35}\,\delta^2-\, \frac{29}{35}\, \G_2\,\right]\ .
\end{align}

\subsection{Renormalization of Quartic Operators}

Finally, we renormalize the quartic operators relevant for the discussion in~\S\ref{sec:RenBias} up to $m=2$. 
The operator $\delta^4$ has already been renormalized in \S\ref{ssec:HOO}, while the Galileon operators are only renormalized by higher-derivative operators. 
The renormalization of $\G_2 \hskip 1pt \delta^2$ follows straightforwardly from the identity
\beq
[\G_2 \hskip 1pt\delta^2]^{{\rm loop}}\ =\ 2\, [\G_2 \hskip 1pt\delta]^{{\rm loop}}\, [\delta]^{{\rm tree}}\ +\ [\G_2]^{{\rm tree}}\,[\delta^2]^{{\rm loop}}\ .
\eeq
We get
\beq
\sum_{\widetilde{\cal O}}Z^{(\G_2\delta^2)}_{\widetilde{\cal O}}\hskip1pt\widetilde{\cal O} \ =\ \frac{8}{3}\sigma^2(\Lambda)\left[\,\delta^2- \frac{3}{8}\G_2\,\right]\ .
\eeq
The remaining operators which need to be renormalized are
$\G_3 \hskip 1pt\delta$, $\Gamma_3\hskip 1pt\delta$ and $[\G_2]^2$.
The counterterms required to cancel the divergences arising from these operators are: 
\begin{align}
\sum_{\widetilde{\cal O}}Z^{([\G_2]^2)}_{\widetilde{\cal O}}\hskip1pt\widetilde{\cal O} &\, =\, -\frac{32}{15} \sigma^2(\Lambda)\left[\,\delta^2 + \frac{1}{4} \G_2\,\right]\ , \\
\sum_{\widetilde{\cal O}}Z^{(\Gamma_3\delta)}_{\widetilde{\cal O}}\hskip1pt\widetilde{\cal O} &\, =\, \frac{64}{105}\sigma^2(\Lambda)\left[\,\delta^2 - \frac{3}{8} \G_2\,\right]\ , \\
\sum_{\widetilde{\cal O}}Z^{(\G_3\delta)}_{\widetilde{\cal O}}\hskip1pt\widetilde{\cal O} &\, =\, -\frac{1}{2}\sigma^2(\Lambda)\,\G_2\ .
\end{align}

\newpage

\section{Non-Renormalization of Galileon Operators}
\label{app:Galileon}

In this appendix, we prove a non-renormalization theorem for the Galileon operators $\G_n$.
In particular, we will show these operators aren't renormalized at leading order in derivatives.
This result is similar to the non-renormalization theorem of Galileons in modified gravity \cite{Hinterbichler:2011tt, deRham:2014zqa}.

\subsection{Non-Renormalization of $\G_2$}

At zeroth order in derivatives, the quadratic Galileon operator  $\G_2 = \nabla_i \nabla_j \Phi \nabla^i \nabla^j \Phi -(\nabla^2 \Phi)^2$ does not get renormalized. Loops arising from $\nabla_i\nabla_j\Phi\nabla^i\nabla^j\Phi$ are exactly canceled by $(\nabla^2\Phi)^2$ and we get
\beq
[\G_2(\Phi)]\ =\ \G_2(\Phi)\, +\, {\cal O}\Big(\frac{\nabla^2}{\Lambda^2}\Big)\ .
\eeq
To prove this result, we only need to show that
\beq
\vev{[\G_2(\Phi)]_\q\hskip 2pt \delta^{(1)}_{\q_1}\cdots \delta^{(1)}_{\q_m}}_{(p){\rm 1PI}}^\prime \ =\ P_1\cdots P_m\left[\, 0\, +\, {\cal O}\Big( \frac{q_i^2}{\Lambda^2}\Big)\, \right]\ .
\eeq
First, let us rewrite $\G_2(\Phi)$ in a more convenient way
\beq
\G_2(\Phi)\, =\, \nabla^i\nabla^j[\Phi{\cal D}_{ij}\Phi]\ , \quad{\rm with\quad}{\cal D}_{ij} \equiv \nabla_{i}\nabla_{j}-\delta^{\mathsmaller{(K)}}_{ij}\nabla^2\ ,
\eeq
which, in Fourier space, becomes
\beq
[\G_2(\Phi)]_{\q}\, =\, -q^{i}q^j\,[\Phi{\cal D}_{ij}\Phi]_\q\ .
\eeq
To complete the proof, we only need to demonstrate that correlation functions of $[\Phi{\cal D}_{ ij}\Phi]_{\q}$ with the linear density contrast $(\delta^{\mathsmaller{(1)}})^m$ are't singular in the limit of vanishing the external momenta~$q_i$. In other words, we need to show that
\beq
\vev{[\Phi{\cal D}_{ ij}\Phi]_{\q}\hskip 2pt\delta^{(1)}_{\q_1}\cdots\delta^{(1)}_{\q_m}}_{(p){\rm 1PI}}^\prime \, =\, P_{1}\cdots P_m + \cdots \ .
\eeq

\subsection*{$\boldsymbol{m=1}$}

We first consider the case $m=1$ and construct a {\it proof by contradiction}.
Let us assume that the correlation function has the following singularity in the limit $q_1\to0$,
\beq
\vev{[\Phi{\cal D}_{ ij}\Phi]_{\q}\hskip 2pt\delta^{(1)}_{\q_1}}_{(p){\rm 1PI}}^\prime \, \xrightarrow{q_1\to0}\, P_1\,\left[\,\frac{a(\Lambda)}{q_1^2}\delta_{ij}\ +\ {\cal O}(q_1^0)\right]\ , \label{equ:XXX}
\eeq
and show that this leads to a contradiction.
The $q_1^{-2}$ divergence in (\ref{equ:XXX}) can only be absorbed if one introduces the potential $\Phi$ as a counterterm
\beq
\sum_{\cal O} Z_{\cal O}^{(\Phi{\cal D}\Phi)}{\cal O}_{ij}\ = \ a(\Lambda)\, \delta_{ij}\,\Phi\, +\, \cdots\ .\label{eq:counter}
\eeq
We now show that this term is not consistent with the symmetries. Indeed, the renormalized operator $[\Phi{\cal D}_{ij}\Phi]$ can be written as
\beq
[\Phi{\cal D}_{ij}\Phi]\ =\ \Phi{\cal D}_{ij}\Phi\ +\ \sum_{\cal O} Z_{\cal O}^{(\Phi{\cal D}\Phi)}\hskip 1pt{\cal O}_{ij}(\Phi,\nabla\Phi,\cdots)\ ,\label{eq:phidphi}
\eeq
where ${\cal O}_{ij}$ are operators which depend locally on $\Phi$ and its derivatives. These operators need not be invariant under a constant shift and a homogeneous boost, since $\Phi{\cal D}_{ij}\Phi$ is not. However, we expect that the renormalized operators satisfy the same symmetries as the bare ones. In particular, shifting the potential~$\Phi$ by a constant $c$ on both sides of (\ref{eq:phidphi}), we get
\beq
[\Phi{\cal D}_{ij}\Phi]\, +\, c\,[{\cal D}_{ij}\Phi]\, =\, \Phi{\cal D}_{ij}\Phi\, +\, c\,{\cal D}_{ij}\Phi \,+\, \sum_{\cal O}Z_{\cal O}^{(\Phi{\cal D}\Phi)}{\cal O}_{ij}(\Phi+c,\nabla\Phi,\cdots)\ .
\eeq
Since the potential is {\it not} renormalized, i.e.~$[{\cal D}_{ij}\Phi]={\cal D}_{ij}\Phi
$, we find
\beq
[\Phi{\cal D}_{ij}\Phi]\, =\, \Phi{\cal D}_{ij}\Phi\, +\, \sum_{\cal O}Z_{\cal O}^{(\Phi{\cal D}\Phi)}{\cal O}_{ij}(\Phi+c,\nabla\Phi,\cdots)\ .\label{eq:phidphis}
\eeq
Comparing eqs.~(\ref{eq:phidphi}) and (\ref{eq:phidphis}), we get
\beq
\sum_{\cal O}Z_{\cal O}^{(\Phi{\cal D}\Phi)}\Big({\cal O}_{ij}(\Phi+c,\nabla\Phi)-{\cal O}_{ij}(\Phi,\nabla\Phi)\Big)\, =\, 0 \ \quad \Rightarrow \ \quad \sum_{\cal O} Z_{\cal O}^{(\Phi{\cal D}\Phi)} \hskip 1pt \frac{\partial{\cal O}_{ij}}{\partial \Phi} \, =\, 0 \ .
\eeq
Since the operators ${\cal O}_{ij}$ form a basis, this is satisfied, if and only if 
\beq
Z_{\cal O}^{(\Phi{\cal D}\Phi)} \hskip 1pt \frac{\partial{\cal O}_{ij}}{\partial \Phi}\, =\, 0\ ,\label{eq:G2shift}
\eeq
for every operator ${\cal O}_{ij}$. Hence, for ${\cal O}_{ij}=\delta_{ij}\Phi$, we have 
\beq
Z_{\Phi}^{(\Phi {\cal D}\Phi)}\, =\, 0\ .\eeq
As a result, the parameter $a(\Lambda)$ in~(\ref{eq:counter}) necessarily vanishes and
\beq
\vev{[\Phi{\cal D}_{ ij}\Phi]_{\q}\,\delta^{(1)}_{\q_1}}^\prime \, \xrightarrow{q_1\to0}\, P_{1}\, +\, \cdots\ .
\eeq

\subsection*{$\boldsymbol{m>1}$}

Let us now consider the general case $m>1$.
We assume that in the limit $|\q_1+\cdots+\q_p|\to0$, with $p\leq m$, the correlation function behaves as
\beq
\vev{[\Phi{\cal D}_{ ij}\Phi]_{\q}\,\delta^{(1)}_{\q_1}\cdots\delta^{(1)}_{\q_m}}_{(p){\rm 1PI}}^\prime \ \xrightarrow{|\q_1+\cdots+\q_p|\to0}\ P_{1}\cdots P_{m} \frac{a(\Lambda)}{|\q_1+\cdots+\q_p|^2}\, \delta_{ij}\ . \label{equ:DIV}
\eeq
 The only counterterm which could remove this divergence is $\delta^{m-p}\hskip 1pt \Phi$.
 By the same logic as before, such a counterterm violates the symmetries of the problem. The divergence in (\ref{equ:DIV}) therefore cannot be present and we have
 \beq
 \vev{[\Phi{\cal D}_{ ij}\Phi]_{\q}\,\delta^{{(1)}}_{\q_1}\cdots\delta^{(1)}_{\q_m}}_{(p){\rm 1PI}}^\prime\, = \, P_1\cdots P_m + \cdots \ .
 \eeq

\subsection{Non-Renormalization of $\G_n$} 

Finally, we prove that the non-renormalization theorem holds for every Galileon operator $\G_n$.  The proof proceeds {\it by induction}. We will assume that
\beq
[\G_{n-1}(\Phi)]\, =\, \G_{n-1}(\Phi)\, +\, {\cal O}\Big(\frac{\nabla^2}{\Lambda^2}\Big)\ , \label{equ:Step1}
\eeq
and prove that this holds for the $n$-th order Galileon operator.

\vskip 10pt
\noindent{\it Definition.---}The $n$-th order Galileon operator can be written as
\beq
\G_n(\Phi)\, \equiv\, -n\,\eta_{(i_1j_1)\,(i_2j_2)\,\cdots\,(i_nj_n)}\nabla^{i_1}\nabla^{j_1}\Phi\,\cdots\nabla^{i_n}\nabla^{j_n}\Phi\ , \label{equ:GnDEF}
\eeq
where the tensor $\eta$ is defined as
\beq
\eta_{(i_1j_1)\,(i_2j_2)\,\cdots\,(i_nj_n)}\, \equiv\, \frac{1}{n!}\sum_{\sigma}(-1)^{\sigma}\delta^{\mathsmaller{(K)}}_{i_1j_{\sigma(1)}}\delta^{\mathsmaller{(K)}}_{i_2j_{\sigma(2)}}\cdots\delta^{\mathsmaller{(K)}}_{i_nj_{\sigma(n)}}\ ,
\eeq
and the sum runs over $n!$ permutations $\sigma$ and $(-1)^\sigma$ represents the signature of the permutation. From the definition, it is clear that $\eta_{(i_1j_1)\,\cdots\,(i_mj_m)\,\cdots\,(i_nj_n)}\, =\, -\eta_{(i_1j_m)\,\cdots\,(i_mj_1)\,\cdots\,(i_nj_n)}$.
Eq.~(\ref{equ:GnDEF}) can therefore be written as 
\beq
{\G_n(\Phi)\, =\, \nabla^{i}\nabla^j\big(\Phi\,T^{(n-1)}_{ij}\big)}\ ,
\eeq
where 
\beq
T_{ij}^{(n-1)}\, \equiv\, -n\,\eta_{(ij)(i_1j_1)\cdots(i_{n-1}j_{n-1})}\nabla^{i_1}\nabla^{j_1}\Phi\cdots\nabla^{i_{n-1}}\nabla^{j_{n-1}}\Phi\ ,
\eeq
or, in Fourier space, 
\beq
[\G_n(\Phi)]_\q\, =\, -q^iq^j[\Phi\,T_{ij}^{(n-1)}]_\q\ .
\eeq

\vskip 4pt
\noindent
{\it Counterterms of $\Phi\hskip1ptT_{ij}^{(n-1)}$.---}We need to prove that the correlation of $[\Phi\hskip1pt T_{ij}^{(n-1)}]_\q$ with $(\delta^{(1)})^m$ is not singular  in the limit where the external momenta, or a partial sum of these external momenta, vanish. To show this, we will prove that symmetries forbid the appearance of counterterms proportional to the operator $\delta_{ij}\Phi$.
As before, we write the renormalized operator as
\beq
[\Phi\hskip 1pt T_{ij}^{(n-1)}]\, =\, \Phi\hskip 1pt T_{ij}^{(n-1)}\, +\, \sum_{\cal O}Z^{(\Phi T)}_{\cal O}{\cal O}_{ij}(\Phi,\nabla\Phi,\cdots)\ ,
\eeq
where again the operators ${\cal O}_{ij}$ are local functions of the potential and its derivatives.
Shifting the potential $\Phi$
 by a constant $c$, we get
\beq
[\Phi\, T_{ij}^{(n-1)}]\, =\, \Phi\, T_{ij}^{(n-1)}\, +\, c\Big(T_{ij}^{(n-1)}-[T_{ij}^{(n-1)}]\Big) \, +\, \sum_{\cal O} Z^{(\Phi T)}_{\cal O}{\cal O}_{ij}(\Phi+c,\nabla\Phi,\cdots)\ ,
\eeq
and, hence,
\beq
 \sum_{\cal O} Z^{(\Phi T)}_{\cal O}\frac{\partial{\cal O}_{ij}}{\partial\Phi}\, =\, [T_{ij}^{(n-1)}]-T_{ij}^{(n-1)} = \sum_{\widetilde{\cal O}}Z^{(T)}_{\widetilde{\cal O}}\,{\widetilde{\cal O}}_{ij}\ ,
\eeq
where $Z^{(T)}_{\widetilde{\cal O}}\hskip 1pt{\widetilde{\cal O}}_{ij}$ are the counterterms required to renormalize $T_{ij}^{(n-1)}$. We see that if there is a counterterm proportional to ${\cal O}_{ij}=\delta_{ij}\Phi$, the renormalization of $T_{ij}^{(n-1)}$ would contain an operator proportional to $\delta_{ij}\mathbb{1}$. As we will show next, this cannot be the case.

\vskip 4pt
\noindent{\it Renormalization of $T^{(n-1)}_{ij}$.---}First, we notice that the trace of $T^{(n-1)}_{ij}$ is proportional to the Galileon operator of order $n-1$,
\beq\delta^{ij}T^{(n-1)}_{ij} \, =\, - \frac{n-5}{n-1}\hskip 2pt \G_{n-1}(\Phi)\ . \label{eq:trace}
\eeq
This strongly constrains the possible counterterms of $T_{ij}^{(n-1)}$.  We write the renormalized operator~$[T_{ij}^{(n-1)}]$ as
\beq
[T_{ij}^{(n-1)}]\ =\ T_{ij}^{(n-1)}\ +\ \sum_{\widetilde{\cal O}} Z^{(T)}_{\widetilde{\cal O}}\hskip 1pt{\widetilde{\cal O}}_{ij}\ .
\eeq
Taking the trace, the operator $T_{ij}^{(n-1)}$  becomes a Galileon operator of order $n-1$ which by (\ref{equ:Step1}) is not renormalized at zeroth order in derivatives. Therefore, the trace of every counterterm which contributes at zeroth order in derivatives has to vanish
\beq
Z^{(T)}_{{\widetilde{\cal O}}^{(0)}} \hskip 1pt \delta^{ij}{\widetilde{\cal O}}^{(0)}_{ij}\, =\, 0\ .
\eeq
As a result, there cannot be a  counterterm proportional to~$\delta_{ij}\mathbb 1$ in $[T_{ij}^{(n-1)}]$ and the counterterms required to renormalize $\Phi\,T_{ij}^{(n-1)}$ do not contain the linear term $\delta_{ij}\Phi$. Consequently, correlations of $\Phi\,T_{ij}^{(n-1)}$  with linear dark matter contrasts are well behaved in the limit of soft momenta
\beq
\vev{[\Phi\,T_{ij}^{(n-1)}]_{\q}\,\delta^{(1)}_{\q_1}\cdots\delta^{(1)}_{\q_m}}_{(p){\rm 1PI}}^\prime \, =\, P_{1}\cdots P_m+ \cdots\ .
\eeq

\newpage
\section{One-Loop Bispectrum}
\label{app:Bispectrum}

In this appendix, we give explicit expressions for the functions ${\cal F}_B =  f_B P_1 P_2$ and ${\cal I}_{B}$, which appear in the one-loop expression of the halo-matter-matter bispectrum 
\beq
\frac{B_{hmm} - b_1^{(R)}B_{mmm}}{P_1 P_2}
= \left(b_2^{(R)}+2b_{\G_2}^{(R)}\big(\mu_{12}^2-1\big) + \sum_{\cal O} b_{{\cal O}}^{(R)} f_B^{[{\cal O}]}\right) +\, \sum_{{\cal O}} b_{{\cal O}}^{(R)}\frac{ {\cal I}^{[{\cal O}]}_B}{P_1 P_2} \ . \label{equ:BI}
\eeq
Operators up to quartic order may contribute to the one-loop bispectrum. 
Classified by their number of legs, these operators can be grouped into three categories
\begin{align}
{\cal O}_{\rm I} &=\{\, \delta\, \}\ ,\\
{\cal O}_{\rm II} &=\{\, [\delta^2]\ ,\ [\G_2]\ ,\ [\Gamma_3]\ ,\ [\Gamma_4]\, \}\ , \label{equ:C2}\\
{\cal O}_{\rm III} &=\{\, [\delta^3]\ ,\ [\G_2\delta]\ , \ [\G_3]\ ,\ [\widetilde\Gamma_4]\ ,\ [\Gamma_3\delta]\,\}\ , \label{equ:C3}
\end{align}
where we have introduced the quartic operators
\begin{align}
\Gamma_4&\equiv\frac{1}{2}\Big(\G_2(\Phi_g)+\G_2(\Phi_v)\Big)-\G_2(\Phi_g,\Phi_v)\ ,\\
\widetilde\Gamma_4&\equiv\G_3(\Phi_g)-\G_3(\Phi_g,\Phi_g,\Phi_v)\ ,
\end{align}
and the third-order Galileon operator
\begin{align}
\G_3(\Phi_\alpha,\Phi_\beta,\Phi_\gamma)&\equiv -\frac{1}{2}\Big[2\nabla_{ij}\Phi_\alpha\nabla^j{}_{k}\Phi_\beta\nabla^{ki}\Phi_\gamma + \nabla^2\Phi_\alpha\nabla^2\Phi_\beta\nabla^2\Phi_\gamma\nonumber\\
&\quad-\big(\nabla_{ij} \Phi_\alpha\nabla^{ij}\Phi_\beta\nabla^2\Phi_\gamma+\nabla_{ij}\Phi_\alpha\nabla^{ij}\Phi_\gamma\nabla^2\Phi_\beta+\nabla_{ij}\Phi_\beta\nabla^{ij}\Phi_\gamma\nabla^2\Phi_\alpha)\Big]\ ,
\end{align}
with $\nabla_{ij} \equiv \nabla_i \nabla_j$ and $(\alpha,\beta,\gamma)\in\{g,v\}$. 
Furthermore, at quartic order, the velocity divergence $\theta \equiv \nabla^2 \Phi_v$ becomes an independent degree of freedom which gives non-trivial contributions to the bispectrum at one-loop. This is captured by the operator~\cite{Chan:2012jj}
\beq
\Delta_4\equiv \left( \nabla^2 \Phi_g - \nabla^2 \Phi_v \right) +\frac{2}{7}\G_2(\Phi_g)-\frac{8}{21}\Gamma_3-\frac{4}{63}\G_3(\Phi_g)\ , \label{equ:D4}
\eeq
which vanishes up to third order.

\subsection{Diagrams}
\label{sec:diag}

The correlation functions involving the operators in (\ref{equ:C2}) and (\ref{equ:C3}) have the following diagrammatic representations: 
\begin{align}
\langle [{\cal O}_{\rm II}]_{\q}\hskip 1pt \delta_{\q_1} \delta_{\q_2} \rangle' \quad &\ \subset\qquad  \ \, \includegraphicsbox[scale=0.7]{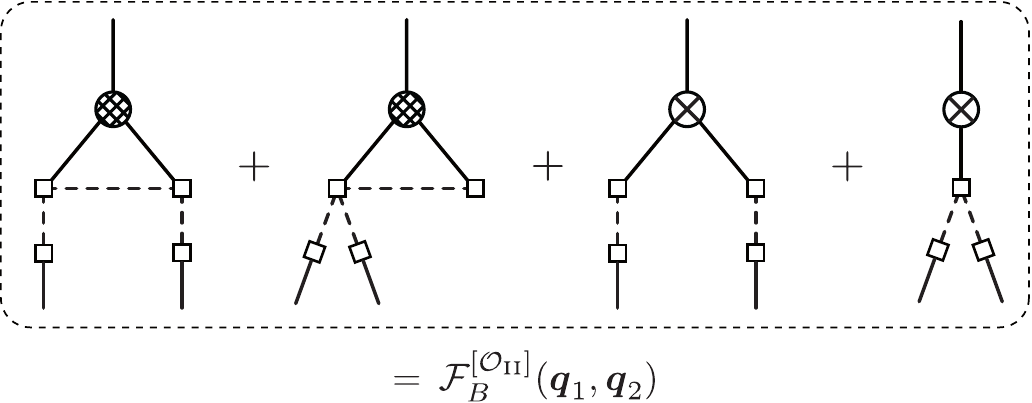} \nonumber\\[3pt] 
&\qquad + \quad \includegraphicsbox[scale=0.7]{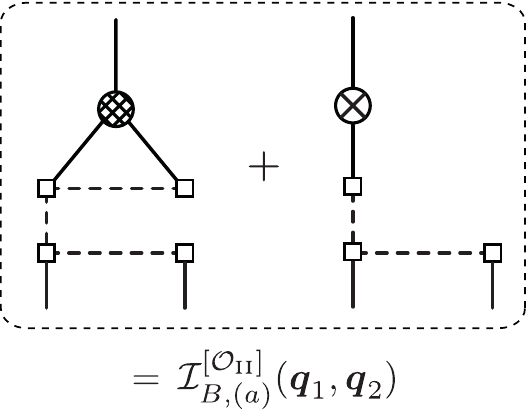} \qquad \nonumber \\[3pt] 
&\qquad + \quad \includegraphicsbox[scale=0.7]{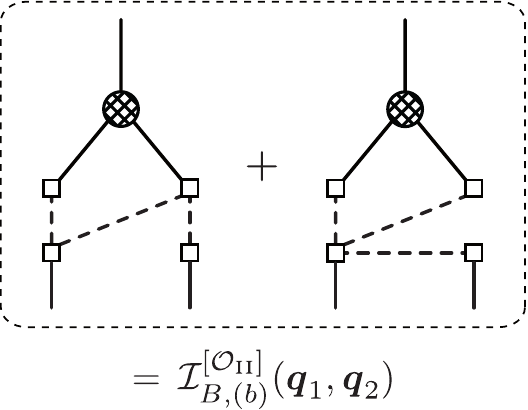} \quad + \quad \includegraphicsbox[scale=0.7]{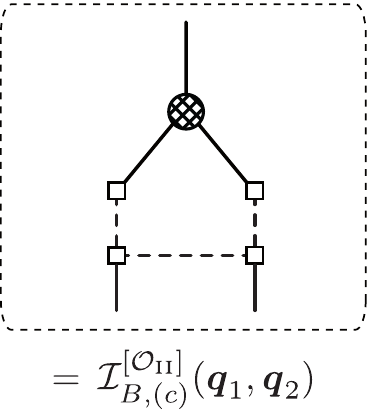}\quad . \label{equ:O2b}\\[10pt]
\langle [{\cal O}_{\rm III}]_{\q}\hskip 1pt \delta_{\q_1} \delta_{\q_2} \rangle' \quad &\ \subset\qquad  \ \, \includegraphicsbox[scale=0.7]{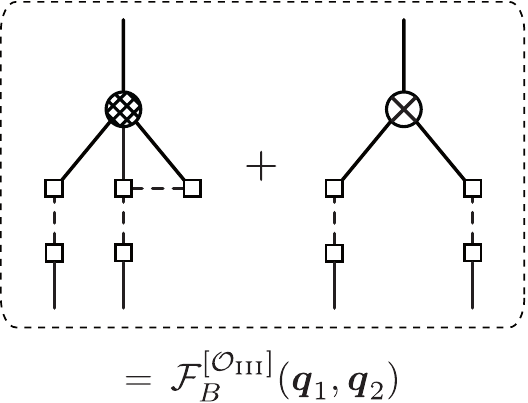} \quad + \quad \includegraphicsbox[scale=0.7]{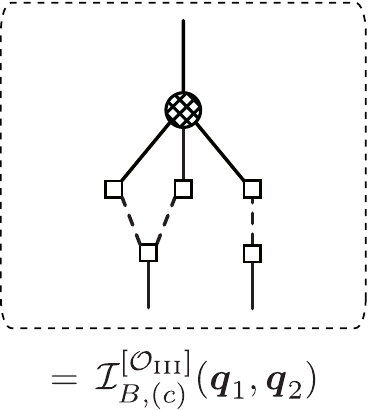} \quad .\label{equ:O3b}
\end{align}
In addition to diagrams similar to the ones in (\ref{equ:O2b}) and (\ref{equ:O3b}),
the operator $\Delta_4$ leads to non-$(p)$1PI diagrams (cf.~fig.~\ref{fig:1PI}):
\begin{align}
\hspace{-3.9cm}\langle [{\Delta}_4]_{\q}\hskip 1pt \delta_{\q_1} \delta_{\q_2} \rangle' \quad &\ \subset\qquad  \ \, \includegraphicsbox[scale=0.7]{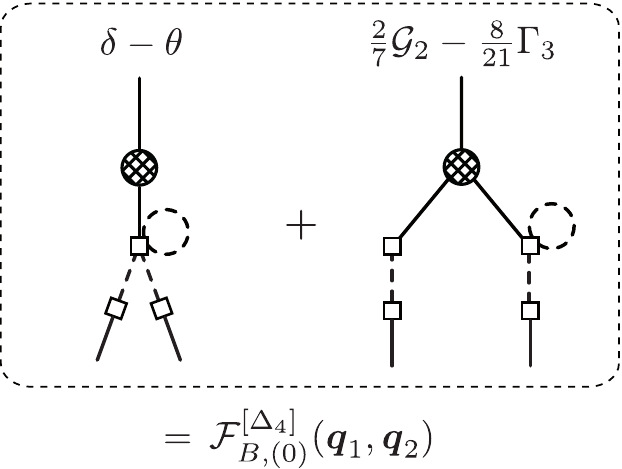}\qquad . \label{equ:D4b}
\end{align}

\subsection{${\cal F}$-terms}

 For ${\cal O}_{\rm II} \equiv \{ [\delta^2] \hskip 1pt, \hskip 1pt[\G_2]\}$  and ${\cal O}_{\rm III} \equiv \{[\delta^3] \hskip 1pt,\hskip 1pt[\G_2\delta]\hskip 1pt,\hskip 1pt [\G_3] \}$, the functions $f_B^{[{\cal O}]}$ in (\ref{equ:BI}) can be written as
\begin{align}
f_{B}^{[{\cal O}_{\rm II}]} & = \int_\p \Big[\hskip 1pt 8\hskip 1pt a_1^{[{\cal O}_{\rm II}]}\hskip 1pt F_2(\q_1,\p)F_2(\q_2,-\p)+12 \hskip 1pt a_2^{[{\cal O}_{\rm II}]}\hskip 1ptF_3(\q_1,\q_2,\p)\Big] P(p) + 2 \hskip 1pt  {\cal C}^{[{\cal O}_{\rm II}]}\ , \label{equ:fb2}\\[4pt]
f_{B}^{[{\cal O}_{\rm III}]} & =\int_\p \Big[\hskip 1pt  2 \hskip 1pt a^{[{\cal O}_{\rm III}]}\hskip 1pt F_2(\q_1,\p)+\{\q_1\leftrightarrow\q_2\}\Big]P(p) + 2\hskip 1pt {\cal C}^{[{\cal O}_{\rm III}]} \ , \label{equ:fb3}
\end{align}
where the functions ${\cal C}^{[{\cal O}_{\rm II}]}$ and ${\cal C}^{[{\cal O}_{\rm III}]}$ are defined in terms of the counterterms of Appendix~\ref{ap:HOO}:
\begin{align}
 {\cal C}^{[{\cal O}]} \equiv Z_{\delta}^{({\cal O})}F_2(\q_1,\q_2) + Z_{\delta^2}^{({\cal O})} + Z_{\G_2}^{({\cal O})}\sigma^2_{\q_1,\q_2}\ , \quad {\rm with}\quad  \sigma^2_{\q_1,\q_2} \equiv  \mu^2_{\q_1,\q_2}-1\ ,
\end{align}
and the functions $a^{[{\cal O}_{\rm II}]}_1$, $a^{[{\cal O}_{\rm II}]}_2$ and $a^{[{\cal O}_{\rm III}]}$  are
\begin{align}
a^{[{\delta^2}]}_1 &= 1\ , \\
 a^{[\delta^2]}_2 &=1\ ,\\
a^{[{\G_2}]}_1 &=\sigma_{\q_1+\p,\q_2-\p}^2\ ,\\
a^{[{\G_2}]}_2 &= \sigma_{\q_1+\q_2+\p,\p}^2\ ,\\
a^{[{\delta^3}]} &= 6\ ,\\
a^{[{\G_2\delta}]} &= 2\Big[\sigma_{\q_1+\p,\p}^2 + \sigma_{\q_2,\p}^2 + \sigma_{\q_1+\p,\q_2}^2\Big]\ ,\\
a^{[{\G_3}]} &=3\Big[  \big(\mu_{\q_1+\p,\p}^2+\mu_{\q_2,\p}^2+\mu_{\q_1+\p,\q_2}^2\big) -  2\, \mu_{\q_1+\p,\p}\, \mu_{\q_2,\p}\, \mu_{\q_1+\p,\q_2} -1\Big]\ .
\end{align}
For  ${\cal O}=\{[\Gamma_3]\hskip1pt,\hskip1pt[\Gamma_4]\hskip 1pt,\hskip 1pt [\widetilde\Gamma_4] \hskip 1pt,\hskip 1pt [\Gamma_3\delta]\}$, we instead have
\begin{align}
f_{B}^{[{\Gamma}_3]}& =4 \int_\p \Big[ 2\hskip 1pt a_1^{[{\Gamma}_3]}\hskip 1pt F_2(\q_1,\p)F_2(\q_2,-\p) +3\hskip 1pt a_2^{[{\Gamma}_3]}\hskip 1pt F_3(\q_1,\q_2,\p)\Big]P(p)  - \{ F_i \to G_i \}\ ,\\[4pt]
f_{B}^{[{\Gamma}_4]}& =4 \int_\p \, \hskip 1pt a_1^{[{\Gamma}_4]}\hskip 1pt\Big[F_2(\q_1,\p)F_2(\q_2,-\p) -  F_2(\q_1,\p)G_2(\q_2,-\p)\Big]P(p) + \{ F_i \leftrightarrow G_i \}\ ,\\[4pt]
f_{B}^{[{\widetilde\Gamma}_4]}& = - \frac{4}{7}  \int_\p \Big[a^{[{\widetilde\Gamma}_4]}\hskip 1pt \sigma_{\q_1,\p}^2 +\{\q_1\leftrightarrow\q_2\}\, \Big]P(p)\ , \\[4pt]
f_{B}^{[{\Gamma}_3\delta]}& =  -\frac{4}{7} \int_\p \Big[ a^{[\Gamma_3 \delta]}\sigma^2_{\q_1,\p}+\sigma^2_{\q_{1} \q_{2},\p}\sigma^2_{\q_1,\q_2}+\{\q_1\leftrightarrow\q_2\}\Big]P(p)+ 2\hskip 1pt {\cal C}^{[\Gamma_3\delta]}\ ,
\end{align}
with $a_{1,2}^{[\Gamma_3]}\equiv a_{1,2}^{[\G_2]}$, $a_1^{[\Gamma_4]}\equiv a_{1}^{[\G_2]}$, $a^{[\widetilde\Gamma_4]}\equiv \tfrac{1}{3}a^{[\G_3]}$ and $a^{[\Gamma_3 \delta]} \equiv 2 (\sigma^2_{\q_1+\p,\q_2}+\sigma^2_{\q_1+\p,\p})$. 
Note that for the first three terms we have not included counterterms in the expression for $f_{B}^{[\cal O]}$ since ${\cal O}=\{[\Gamma_3]\hskip 1pt,\hskip 1pt[\Gamma_4] \hskip 1pt,\hskip 1pt [\widetilde\Gamma_4]\}$ are Galileon operators which are not renormalized at leading order in derivatives.

Finally, we consider the operator $\Delta_4$ of eq.~(\ref{equ:D4}). In this case, we find
\begin{align}
f_{B}^{[\Delta_4]} &=\frac{2}{7}f_{B}^{[\G_2]}-\frac{8}{21}f_{B}^{[\Gamma_3]}-\frac{4}{63}f_{B}^{[\G_3]} + f_{B,({\rm 0})}^{[\Delta_4]} \ ,
\end{align}
where the last term corresponds to the non-$(p)$1PI diagrams in eq.~(\ref{equ:D4b}): 
\begin{align}
f_{B,({\rm 0})}^{[\Delta_4]}&\ =\ 
12 \int_\p\Big[\, \big(F_4(\q_1,\q_2,\p,-\p)-G_4(\q_1,\q_2,\p,-\p)\big)\nonumber\\
&\hskip 50pt +\frac{1}{7} \hskip 1pt \sigma_{\q_1,\q_2}^2 \big(F_3(\q_1,\p,-\p)+F_3(\q_2,\p,-\p)\big)\nonumber\\
&\hskip 50pt  -\frac{4}{21} \hskip 1pt  \sigma_{\q_1,\q_2}^2 \big(F_3(\q_1,\p,-\p)-G_3(\q_1,\p,-\p)+\{\q_1\leftrightarrow \q_2\} \big) \Big]P(p)\ .
\end{align}

\subsection{${\cal I}$-terms}
The functions ${\cal I}^{[{\cal O}]}_{B}$ can be written as the sum of three terms corresponding to the three types of diagrams in (\ref{equ:O2b}) and (\ref{equ:O3b}):
\beq
{\cal I}_{B}^{[{\cal O}]} = {\cal I}_{B, (a)}^{[{\cal O}]}+  {\cal I}_{B, (b)}^{[{\cal O}]}+ {\cal I}_{B, (c)}^{[{\cal O}]} \ .
\eeq

\begin{itemize}
\item The first term is non-zero only for the operators ${\cal O}_{\rm II} = \{[\G_2]\hskip 1pt,\hskip 1pt[\Gamma_3]\}$:
\begin{align}
 {\cal I}_{B,(a)}^{[{\cal O}_{\rm II}]} &\equiv 2\hskip 1pt  P_1  \, {\cal F}^{[{\cal O}_{\rm II}]}(|\q_1+\q_2|) F_2(\q_1+\q_2,-\q_1) +\{\q_1\leftrightarrow \q_2\}\ , 
 \end{align}
 where ${\cal F}^{[{\cal O}_{\rm II}]}$ was defined in~(\ref{equ:FG2}) and (\ref{equ:FGamma3}).

 \item The second term is
 \begin{align}
 {\cal I}_{B, (b)}^{[{\cal O}]} & \equiv P_1\int_\p i^{[{\cal O}]}_{(b)}\, 
P(p)P(|\q_2-\p|) \,+\, \{\q_1\leftrightarrow\q_2\} \ .
\end{align}
For the operators  ${\cal O}_{\rm II} = \{ [\delta^2]\hskip 1pt,\hskip 1pt [\G_2]\hskip 1pt,\hskip 1pt [\Gamma_3]\}$, the functions
 $i^{[{\cal O}_{\rm II}]}_{(b)}$ are
\begin{align}
i^{[{\delta}^2]}_{(b)} &= 8 \hskip 1pt F_2(\q_2-\p,\p)  F_2(-\p,-\q_1) + 6\hskip 1pt   F_3(-\q_1,\p,\q_{12}-\p)\ ,  \\[4pt]
i^{[{\G_2}]}_{(b)} &=8\hskip 1pt  \sigma_{\q_2-\p,\q_1+\p}^2 F_2(\q_2-\p,\p)F_2(-\p,-\q_1)  + 6\hskip 1pt  \sigma_{\q_{12}-\p,\p}^2F_3(-\q_1,\p,\q_{12}-\p)  \ , \\[4pt]
i^{[{\Gamma_3}]}_{(b)} &=8 \hskip 1pt \sigma_{\q_2-\p,\q_1+\p}^2 F_2(\q_2-\p,\p)F_2(-\p,-\q_1)  - \{ F_i \to G_i \}\ .
\end{align}
For the operators  ${\cal O}_{\rm III} = \{ [\delta^3]\hskip 1pt,\hskip 1pt [\G_2 \delta]\hskip 1pt,\hskip 1pt[\G_3]\}$, the functions
 $i^{[{\cal O}_{\rm III}]}_{(b)}$ are 
\begin{align}
i^{[{\delta}^3]}_{(b)} &= 6\hskip 1pt  F_2(\q_2-\p,\p)\ ,  \\
i^{[{\G}_2\delta]}_{(b)} &= 2 \Big[\sigma_{\q_2-\p,\p}^2 +2\sigma_{\q_1,\p}^2\Big] F_2(\q_2-\p,\p)\ , \\
i^{[{\G}_3]}_{(b)} &=  3 \Big[ \big(\mu_{\q_2-\p,\p}^2+2\mu_{\q_1,\p}^2\big)- 2 \mu_{\q_2-\p,\p}\, \mu_{\q_2-\p,\q_1}\, \mu_{\q_1,\p} -1\Big] F_2(\q_2-\p,\p)\ . 
\end{align}

\item Finally, the third term in~(\ref{equ:O2b}) only gets contributions from ${\cal O}_{\rm II} = \{[\delta^2]\hskip 1pt,\hskip 1pt[\G_2] \}$:
\begin{align}
{\cal I}_{B, (c)}^{[{\cal O}_{\rm II}]}& = 4\int_\p i^{[{\cal O}_{\rm II}]}_{(c)}\, F_2(\q_1+\p,-\p)F_2(\p,\q_2-\p) \, P(|\q_1+\p|)P(|\q_2-\p|)P(p)\ ,
\end{align}
where
\begin{align}
i^{[{\delta^2}]}_{(c)} &= 1\ ,\\
i^{[{\G_2}]}_{(c)} &= \sigma_{\q_1+\p,\q_2-\p}^2\ .
\end{align}
\end{itemize}

\newpage
\addcontentsline{toc}{section}{References}
\bibliographystyle{utphys}
\bibliography{BiasRenorm,books}

\end{document}